\documentclass{jaa}
\usepackage{natbib}

\usepackage{graphicx}
\usepackage{dcolumn}
\usepackage{bm}

\usepackage{aasmacros}
\usepackage[table,xcdraw,svgnames]{xcolor}
\usepackage[colorlinks]{hyperref}
\hypersetup{
    citecolor=Blue,
    linkcolor=Red,   
    urlcolor=Blue}

\graphicspath{{./plots/}}

\begin{document}\sloppy

\title{Gravitational Lensing of Core Collapse Supernova Gravitational Wave Signals}

\author{RAHUL RAMESH, ASHISH KUMAR MEENA and JASJEET SINGH BAGLA\textsuperscript{*}.}
\affilOne{Department of Physical Sciences, IISER Mohali, Sector 81, SAS Nagar, Punjab, 
India - 140306.}

\twocolumn[{

\maketitle

\corres{jasjeet@iisermohali.ac.in}

\msinfo{xx yy 2021}{xx yy 2021}

\begin{abstract}
We discuss the prospects of gravitational lensing of gravitational 
waves (GWs) coming from core-collapse supernovae (CCSN). 
As the CCSN GW signal can only be detected from within our own
Galaxy and the local group by current and upcoming ground-based GW 
detectors, we focus on microlensing. 
We introduce a new technique based on analysis of the power spectrum
and association of peaks of the power spectrum with the peaks of the
amplification factor to identify lensed signals. 
We validate our method by applying it on the CCSN-like mock signals 
lensed by a point mass lens.
We find that the lensed and unlensed signal can be differentiated
using the association of peaks by more than one sigma for lens masses
M$_{\rm L} {>} 150{\rm M}_{\odot}$.  
We also study the correlation integral between the power spectra and
corresponding amplification factor. 
This statistical approach is able to differentiate between unlensed
and lensed signals for lenses as small as M$_{\rm L} {\sim} 15{\rm
  M}_{\odot}$.
Further, we demonstrate that this method can be used to estimate
the mass of a lens in case the signal is lensed.
The power spectrum based analysis is general and can be applied to any
broad band signal and is especially useful for incoherent signals. 
\end{abstract}

\keywords{Gravitational lensing: micro; gravitational waves}

}]

\doinum{}
\artcitid{}
\volnum{000}
\year{0000}
\pgrange{1--}
\setcounter{page}{1}
\lp{1}

\section{Introduction}
\label{sec:introduction}

The detection of gravitational waves (GWs) from merging binaries 
by Laser Interferometer Gravitational-wave Observatory (LIGO) and 
Virgo \citep[][]{2019PhRvX...9c1040A, 2020arXiv201014527A}
marked the beginning of a new field to study the Universe. 
With improvements in sensitivity of existing detectors and 
upcoming facilities like the Kamioka Gravitational Wave 
Detector \citep[KAGRA,][]{2012CQGra..29l4007S}, and LIGO-India 
\citep{2013IJMPD..2241010U}, the number of such GW sources will 
continuously increase.

In addition to chirp signals emitted by merging binaries, 
core-collapse supernovae (CCSN) also emit GWs in LIGO/Virgo 
frequency range ($10$~Hz -- $10^4$~Hz) \citep{2009CQGra..26f3001O, 
2011LRR....14....1F}.
Detailed multi-dimensional CCSN simulations and other studies \citep[e.g.,][]
{2013ApJ...768..115O, 2014PhRvD..89d4011K, 2019ApJ...876L...9R,
  2021arXiv211003131S} 
indicate that the corresponding GW signals are around $\sim$~1 second
long and can  
only be detected up to a distance of $100$~kpc with the existing 
and upcoming advanced network of ground-based detectors.
This distance covers our galaxy and the nearby Magellanic Clouds, 
which leads to an event rate of $\sim$~2 CCSN/100 yr.
However, some extreme models also suggest the possibility of GW
signal detection from a distance of $\sim$~100 Mpc by the
advanced ground-based network with an event rate of $\sim$~2 CCSN/yr  
\citep[e.g.,][]{2002ApJ...565..430F, 2006PhRvL..96t1102O, 
2007ApJ...658.1173P}.  However, these are already constrained by a
lack of observations of such events in the last few years.

As discussed (extensively) in literature, GW signals are also 
subject to gravitational lensing similar to electromagnetic 
radiation 
\citep[e.g.,][]{2018arXiv180205273B, 2018arXiv180707062H,
2018MNRAS.476.2220L, 2021arXiv210514390X}. 
However, due to the large wavelength of GWs in the LIGO/Virgo 
frequency range, microlensing can introduce frequency-dependent 
effects in the GW signal
\citep[e.g.,][]{2018PhRvD..98j3022C, 2019A&A...627A.130D, 
2020MNRAS.492.1127M, 2021arXiv210203946M}.
Currently, LIGO/Virgo can detect GWs emitted by merging 
binaries out to cosmological distances. Hence, one can look for 
the signature of both strong lensing and microlensing in these 
signals. 
However, as mentioned above, due to an order of magnitude 
weaker GW signal in CCSN, one can only detect them from within 
the local group.
Hence, only galactic microlenses can introduce the lensing effect 
in CCSN GW signals.

In our current work, we look at the prospects of observing 
galactic microlensing signatures in the CCSN GW signals.
We also discuss possible methods based on the power spectrum
analysis of the GW signal that can be used to find lensed signals.
We validate our methodology using CCSN-like mock GW signals generated
using the Gaussian random fields. 
Although this method is able to differentiate between lensed
and unlensed GW signals, it is not effective for all lens
masses. Hence, we also discuss the possible limitations of
our method.

This paper is organized as follows:
In \S\ref{sec:basic_lensing}, we review the relevant
basics of gravitational lensing. 
In \S\ref{sec:gaussian}, we consider mock waveforms as model 
signals to give an overview of the various features that are 
expected when signals are lensed in the regime where wave effects 
are non negligible.
In \S\ref{sec:sig_gen}, we describe the procedure that 
we have used to generate CCSN-like GW signals.
We present results in \S\ref{sec:results}.
In \S\ref{ssec:ccsn_lensed}, we discuss the effect 
of lensing on the the CCSN-like GW signal and the corresponding
power spectrum.  
The methods to identify lensed signals are discussed in 
\S\ref{ssec:lensed_identify}. In \S\ref{ssec:constraint}, we discuss a
possible application of the correlation integral in the estimation of
(point mass) lens parameters. 
Limitations of these methods are discussed in \S\ref{ssec:limitations}.
Conclusion and summary are presented in \S\ref{sec:conclusions}.

\begin{figure*}[ht!]
	\centering
	\includegraphics[height=10cm, width=16cm]{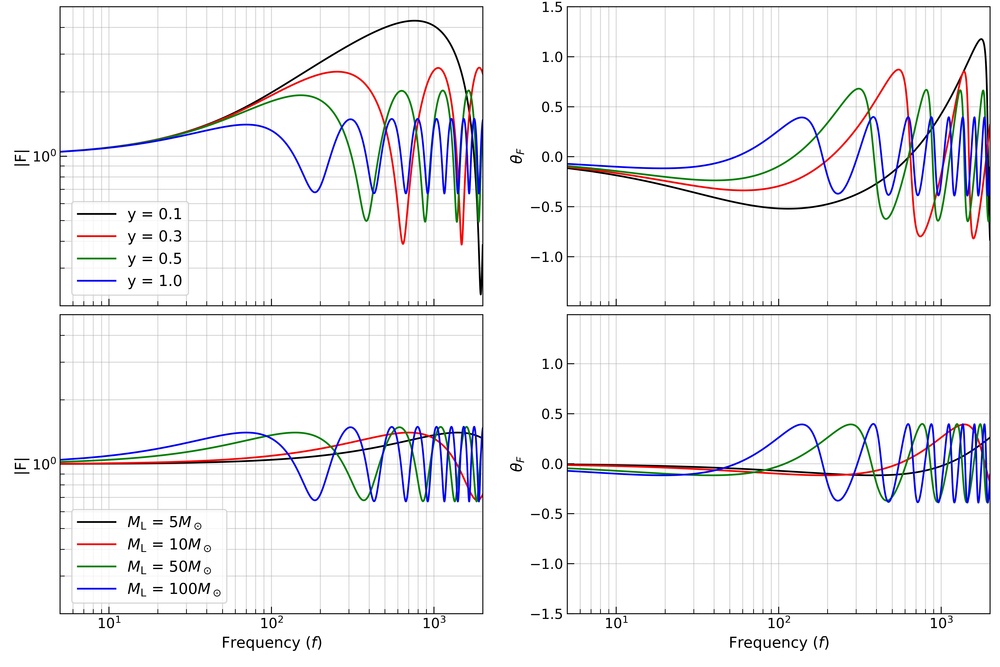}
	\caption{Gravitational lensing due to a point mass lens: 
	The top row represents the amplification factor ($|F|$) 
	and phase shift ($\theta_F$) as a function of frequency 
	due to a point mass lens of 
	100M$_\odot$ with different source positions ($y$) in 
	left and right panel, respectively.The lens redshift has 
	been set to zero (i.e., $z_{\rm L} = 0$). Similarly, the bottom 
	row shows the amplification factor and phase shift factor 
	due to a point lens with different lens mass values with the
	source position fixed at $y=1$.}
	\label{fig:point_lens}
\end{figure*}

\section{Basics of Gravitational Lensing}
\label{sec:basic_lensing}

For cases where the wavelength of the signal coming from a source 
is very small compared to the Schwarzschild radius of the lens 
($\lambda \ll {\rm R}_{\rm Sch}$), one can use the conventional 
geometric optics to study the gravitational lensing \citep[][]
{1992grle.book.....S}. 
However, if the signal wavelength is of the order of the 
Schwarzschild radius of the lens ($\lambda \sim {\rm R}_{\rm Sch}$) 
then wave effects become non-negligible \citep[][]
{1999PThPS.133..137N, 2003ApJ...595.1039T}.
In such cases, the amplification factor is given as:
\begin{equation}
    F\left(f,\mathbf{y}\right)=\frac{\xi_0^2}{c}
    \frac{D_{\rm s} \left(1+z_{\rm d}\right)}{D_{\rm d} D_{\rm ds}}
    \frac{f}{\textit{i}} \int d^2\mathbf{x} \:
    \exp\left[2\pi\textit{i} f t_{\rm
        d}\left(\mathbf{x},\mathbf{y}\right)\right],  
\label{eq:amplification_factor}
\end{equation}
where $f$ is the frequency of the signal and $z_{\rm d}$ is the 
lens redshift.
$D_{\rm d}$, $D_{\rm s}$, and $D_{\rm ds}$ are the angular diameter 
distances from observer to lens, observer to source, and from lens 
to source, respectively.
$\mathbf{y} = \boldsymbol{\eta}D_{\rm d}/\xi_0 D_{\rm s}$ and 
$\mathbf{x} = \boldsymbol{\xi}/\xi_0$ are the dimensionless source 
and image position in the source and image plane, respectively.
Here $\xi_0$ is an arbitrary length scale to make the equation 
dimensionless.
The time-delay factor, $t_{\rm d}$, between lensed and unlensed 
images is given as:
\begin{equation}
    t_{\rm d} = \frac{\xi^2_0}{c}\frac{D_{\rm s}}{D_{\rm d} D_{\rm ds}}(1+z_{\rm d})
    \left[\frac{(\bm{x}-\bm{y})^2}{2}-\psi(\bm{x})+\phi_m(\bm{y})\right],
    \label{eq:time_delay}
\end{equation}
where $\psi(\bm{x})$ is the lens potential and $\phi_m(\bm{y})$ 
is a constant which can be chosen to simplify the calculations. 
In our present work, we choose $\phi_m(\bm{y})$ such that the time 
delay corresponding to the global minimum is zero.

One cannot solve Equation \ref{eq:amplification_factor} 
analytically for complex lens mass models. 
Only the isolated point mass lens (of mass ${\rm M}_{\rm L}$) has an 
analytic solution which is given as \citep[][]{1974PhRvD...9.2207P}:
\begin{eqnarray}
    F\left(\omega,y\right) &=& \exp\left[\frac{\pi
                           \omega}{4}+\frac{\textit{i}\omega}{2}
                           \Bigg\{\ln\left(\frac{\omega}{2}\right) -
                           2\phi_m\left(x_m\right)\Bigg\} 
                           \right]\nonumber\\  
&&
   \Gamma\left(1-\frac{\textit{i}\omega}{2}\right){}_{1}F_{1}
   \left(\frac{\textit{i}\omega}{2},1;\frac{\textit{i}\omega }{2}y^2
   \right),  
\label{eq:point_wave}
\end{eqnarray}
where
\begin{equation}
\omega = \frac{8\pi G {\rm M}_{\rm L} (1+z_{\rm d})f}{c^3}; \quad
\phi_m(y) = \frac{(x_m-y)^2}{2} - \ln x_m,
\end{equation}
with $x_m = \left(y+\sqrt{y^2+4}\right)/2$.
Here, Einstein radius of the point mass lens is used as the 
length scale.  
In case of geometric optics limit, $\lambda \ll 
{\rm R}_{\rm Sch}$, Equation \ref{eq:point_wave} can be written
as
\begin{equation}
	F(f,y) = \sqrt{\mu_+} - \iota \sqrt{\mu_-} \exp[2\pi \iota f
          \Delta t_{\rm d}], 
	\label{eq:point_geo}
\end{equation}
where $\mu_\pm$ denote the amplification factor for primary and 
secondary image for a point mass lens in geometric optics limit 
and  $\Delta t_{\rm d}$ is the time delay between these two images. 
In reality, stellar mass microlenses are extended objects. 
Hence, one needs to verify that the point mass lens approximation is valid
for microlenses.
We validate this approximation in \ref{sec:appendix_A}, and consider
microlenses as point masses throughout this work.

\begin{figure*}[ht!]
	\centering
	\includegraphics[height=6cm, width=16cm]{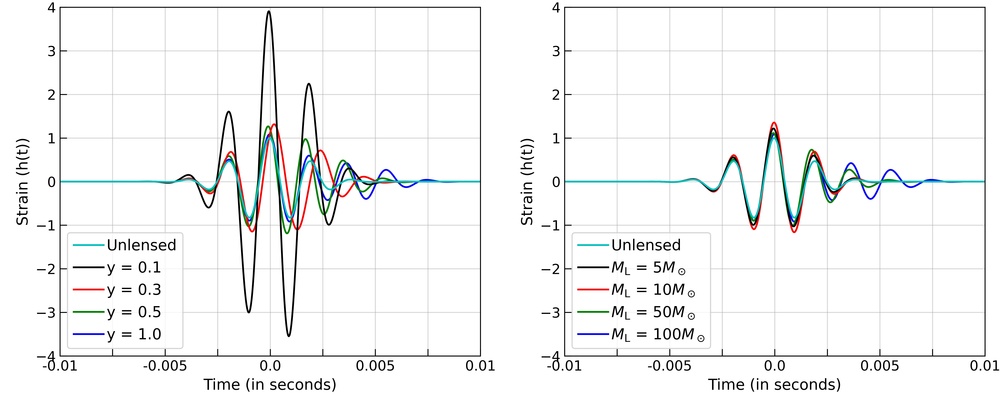}
	\caption{Lensing of a Gaussian wave packet due to a point mass
	lens: The left panel represents the unlensed (cyan) and lensed
	Gaussian wave packet with a lens mass of $100{\rm M}_\odot$
	and different source positions ($y$). The $y$-values for different
	lensed curves are shown in the legend. Similarly, the right panel
	represents the unlensed (cyan) and lensed Gaussian wave packets
	with different lens mass values and fixed source position ($y=1$).
	In both panels, the x-axis represents the time and y-axis represents
	the normalized strain of the GW signal.
	As the source position and lens mass values are same as the Figure
	\ref{fig:point_lens}, one can clearly see the same trend in both
	figures.}
	\label{fig:gaussian}
\end{figure*}

\begin{figure*}[ht!]
	\centering
	\includegraphics[height=6cm, width=16cm]{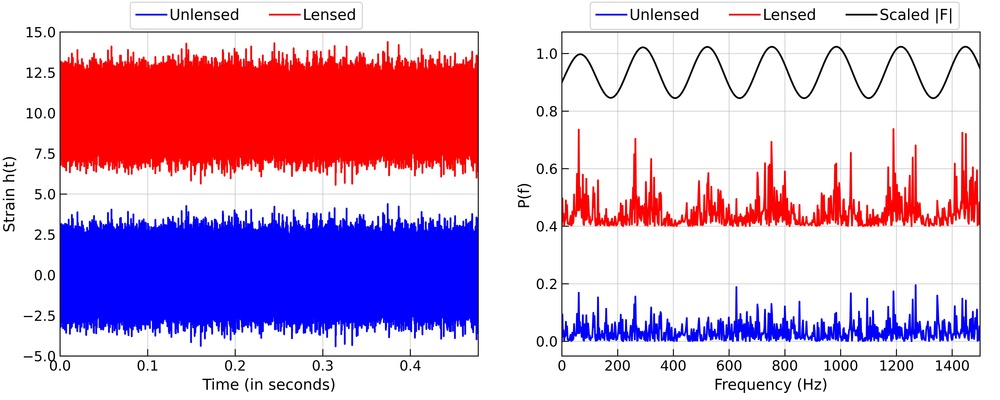}
	\caption{Lensing of white noise due to a point mass lens: The 
	left panel shows the unlensed (blue) and lensed (red) signals. 
	We here consider ${\rm M}_{\rm L} = 100 {\rm M}_\odot$. The two signals have been 
	vertically displaced for better visibility. As can be seen, it 
	is almost impossible to differentiate between the two signals, 
	owing to the fact that white noise shows no particular evolution 
	of frequency with time. The right panel shows the corresponding 
	power spectra, in corresponding colors. Again, the two curves have 
	been vertically displaced. For reference, we also show (a scaled 
	version of) the modulus of the amplification factor. The power of 
	the lensed signal oscillates in accordance with the oscillations 
	of the amplification factor, while the power spectrum of the 
	unlensed signal shows stochastic oscillations.}
	\label{fig:white_noise}
\end{figure*}

From Equation \ref{eq:point_wave}, for a given frequency value, one
  can see that lensing due 
to a point mass lens depends only on the lens mass and
the source position.
Figure~\ref{fig:point_lens} represents the dependency of 
amplification factor on these two parameters.
In the top row, the lens mass is fixed at 100M$_\odot$ and the 
source position is varied. 
One can see that, as the source moves towards the center, the
amplitude of oscillations in both amplification factor($|F|$; left
panel) and phase shift ($\theta_{\rm F}$; right panel) increases. 
However, the frequency of these oscillations decreases.
Apart from that, the first peak in the amplification factor 
($|F|$) also moves towards the high frequencies. 
On the other hand, the lower row represents the variation in the
lens mass for a fixed source position ($y=1$).
One can see that fixing the source position determines the 
amplitude of the oscillations in both the amplification factor ($|F|$)
and the phase shift ($\theta_{\rm F}$).
An increase in the lens mass increases the frequency of the 
oscillations starting from the lower frequency values.

Throughout this paper, unless otherwise mentioned, we always fix
the source at $y=1$. 

\begin{figure*}[h!]
	\centering
	\includegraphics[height=6cm, width=16cm]{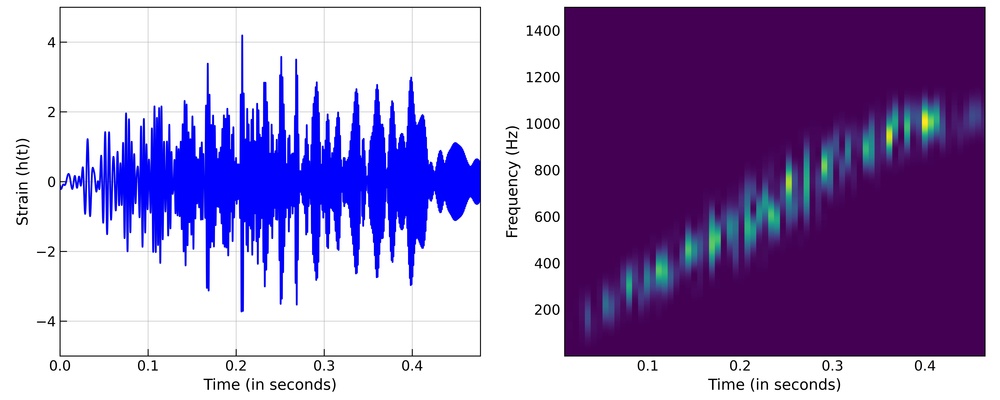}
	\caption{Simulated CCSN-like GW signal: The left panel represents the
	CCSN-like GW signal generated by the method described in 
	\S\ref{sec:sig_gen}. The x-axis represents the time and y-axis 
	represents the normalized strain. The right panel represents the
	corresponding frequency-time spectrogram. The x-axis represents
	the time and the y-axis represent the frequency values.}
	\label{fig:ccsn_gw}
\end{figure*}

\begin{figure*}[ht!]
	\centering
	\includegraphics[height=6cm, width=16cm]{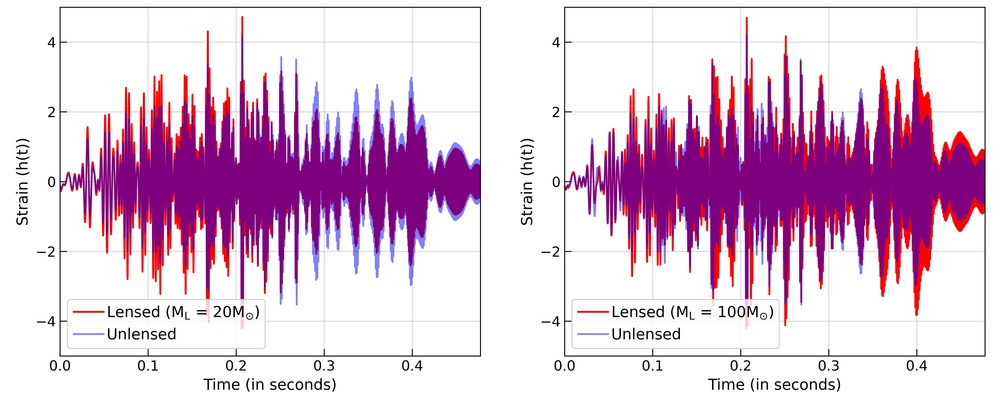}
	
	\vspace{0.5cm}

	\includegraphics[height=6cm, width=16cm]{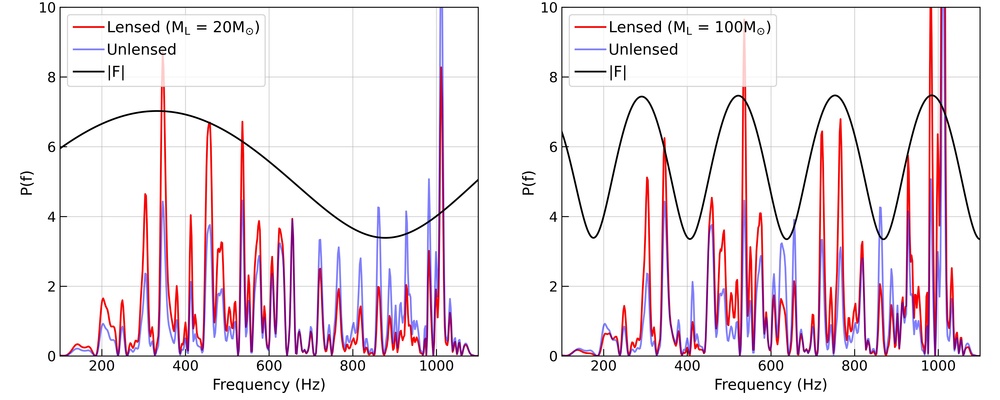}
	\caption{Lensed CCSN-like GW signal and the corresponding Power spectrum
	corresponding to Figure	\ref{fig:ccsn_gw}: The top left 
	panel represents the lensed (unlensed) signal in red (blue) with
	the lens mass 20M$_{\odot}$. The top right panel represents the
	same signal lensed with a lens mass of 100M$_{\odot}$.
	The bottom left panel represents the power spectrum (P(f)) of 
	lensed (by a 20M$_{\odot}$) and unlensed signal in red and blue, 
	respectively. The black solid line represents the amplification 
	factor ($|F|$) due to lens mass. The amplification factor is 
	scaled by a factor of five to (i.e., 5$|F|$) to separate 
	it from the power spectrum. The right panel is a
	similar plot for the lens mass of 100M$_{\odot}$.}
	\label{fig:ccsn_gw_ps}
\end{figure*}

\begin{figure*}[ht!]
	\centering
	\includegraphics[height=10cm, width=16cm]{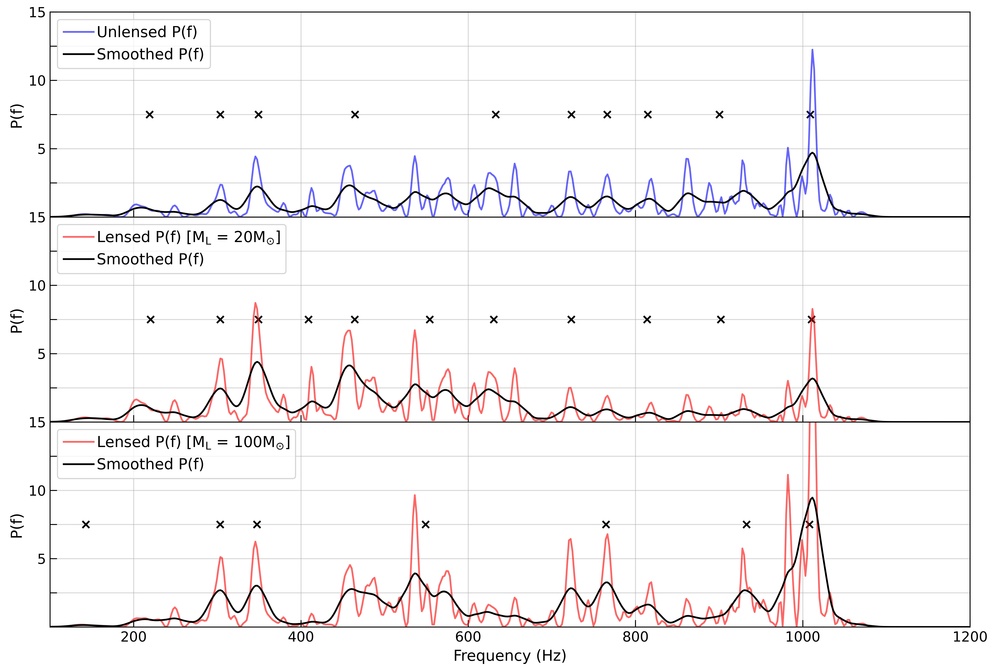}
	\caption{Finding peaks in the power spectrum: In the top
	panel, the blue line represents power spectrum (P(f)) of 
	unlensed signal from Figure \ref{fig:ccsn_gw}. The black 
	line represents the smoothed power 
	spectrum using the Savitzky-Golay filter. The `$\times$' 
	points represent the frequency values corresponding to peaks 
	estimated using the algorithm mentioned in the main text. Similarly, 
	in the middle and bottom panel,
	the red line represents the lensed power spectrum 
	corresponding to a lens mass of 20M$_{\odot}$ and 
	100M$_{\odot}$, respectively.}
	\label{fig:ccsn_gw_ps_smooth}
\end{figure*}

\section{Lensing of Ideal Gravitational Wave Signals}
\label{sec:gaussian}

Unlike chirp signals emitted from a merging binary, CCSN events 
give rise to broadband incoherent GW signals with a duration of
${\sim}1$ second in the time domain.
Due to its simplicity and broadband nature, a Gaussian wave 
packet is an obvious starting point to gain initial insights about 
the lensing of CCSN GW signals.
The Gaussian wave packet (in time domain) is given as 
\begin{equation}
	h(t) = \exp\left[-2\pi^2 \sigma^2(t-t_0)^2\right]
	\exp(2\pi \iota f_0 t),
	\label{eq:wavepacket}
\end{equation}
where $f_0$ and $\sigma$ are the central frequency and standard
deviation of the wave packet (in frequency domain).
Figure \ref{fig:gaussian} shows the lensing effect due to a point 
mass lens on a Gaussian wave packet with a central frequency of 
500 Hz and a standard deviation of 100 Hz.
In both panels, the x-axis represents the time, and the y-axis 
shows the signal amplitude. 
Similar to Figure \ref{fig:point_lens}, the left panel of the 
source position has been varied while fixing the lens mass to 
$100{\rm M}_\odot$, whereas in the right panel, the lens mass 
is varied while keeping the source at $y=1$.
In the right panel, one can see that both lensed and unlensed 
signals arrive at the same time as we set the time delay for 
the minima image to zero.  
However, the lensed signal is not only amplified but also has a 
spread in time larger than the unlensed signal. It can be understood 
from the right panel of the figure: one can see that as we increase the lens 
mass, the duration of the signal increases. 
This is because the lensed signal represents the interference 
between the minima and saddle images which forms in the 
geometric optics regime. 
As we increase the lens mass, the time delay between 
these two image increases, so the final signal has a longer duration 
as the saddle image part gets delayed as compared to the minima. 
It becomes clearer if we increase the lens mass (or move 
the source away from the center), which increases the time delay 
between the two images. 
Eventually, the time delay between the two images exceeds 
the duration of the signal, and we observe two separate signals. 
Equivalent to the above description, one can also think of 
lensing in terms of oscillations of the amplification factor: 
as the amplification factor oscillates faster (corresponding to 
a greater time delay), a greater distortion is observed in the signal. 

\begin{figure*}[ht!]
	\centering
	\includegraphics[height=7.5cm, width=16cm]{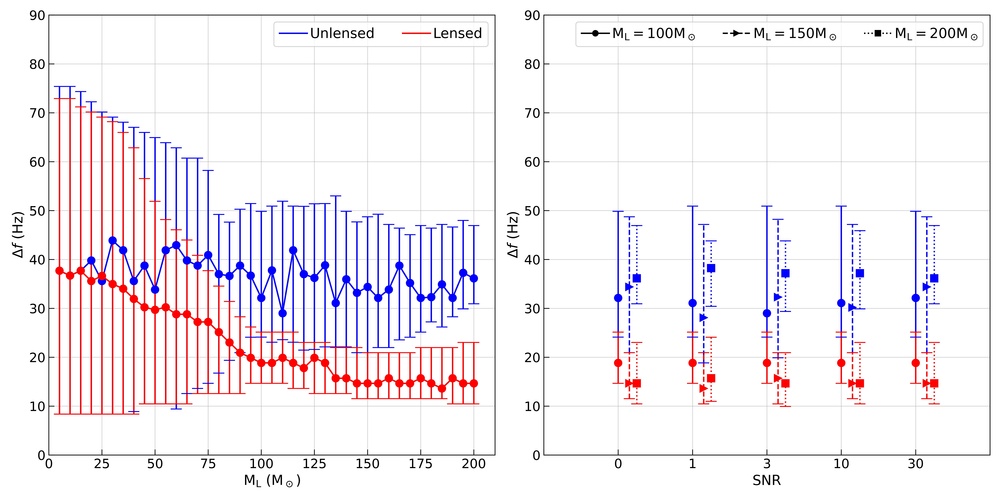}
	\caption{Results of the peak matching method: The left panel 
	represents the efficiency of the peak matching method in the
	absence of external noise. As elaborated in the main text, 
	$\Delta f$ is the average distance between peaks of the 
	amplification factor and estimated `genuine-peaks' of the power 
	spectrum. Blue (red) solid points are median values corresponding 
	to unlensed (lensed) signals, and errors bars correspond to 
	$16^{\rm th}$ and $84^{\rm th}$ percentiles. 
	The right panel represents the effect of SNR on the median values
	and error bars for three different mass values as mentioned in the
	plot legend. A SNR of zero does not mean that there is no signal, 
	but instead refers to the absence of external noise.
	The various points for different lens masses for a particular
	SNR value are horizontally shifted for better visualization.
	One can see that the external noise does not significantly affect the final results.}
	\label{fig:pk_match_SNR}
\end{figure*}

\begin{figure*}[ht!]
	\centering
	\includegraphics[height=7.5cm, width=16cm]{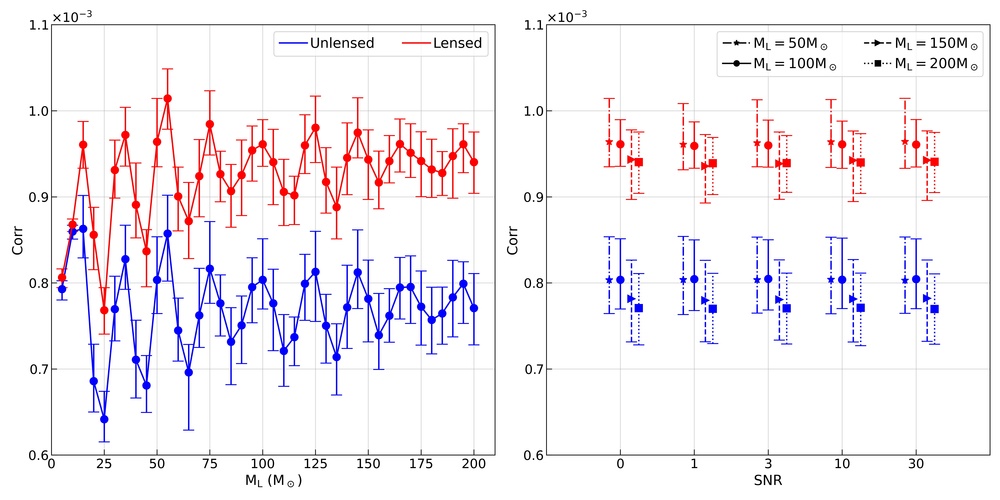}
	\caption{Correlation Integral computed at various lens 
	masses for different cases with varying SNR. Similar to 
	Figure \ref{fig:pk_match_SNR}, blue (red) solid points are 
	median values corresponding to the unlensed (lensed) signals, 
	and errors bars represent the $16^{\rm{th}}$ and $84^{\rm{th}}$ 
	percentile values. The left panel represents the correlation 
	values in the absence of external noise. The effect of external noise
	is presented in the right panel for three different lens mass values. 
	Similar to the peak matching algorithm, the external noise does not play
	a significant role in the correlation integral values.}
	\label{fig:corr_intgl}
\end{figure*}

\begin{figure*}
	\centering
	\includegraphics[height=5.3cm, width=14cm]{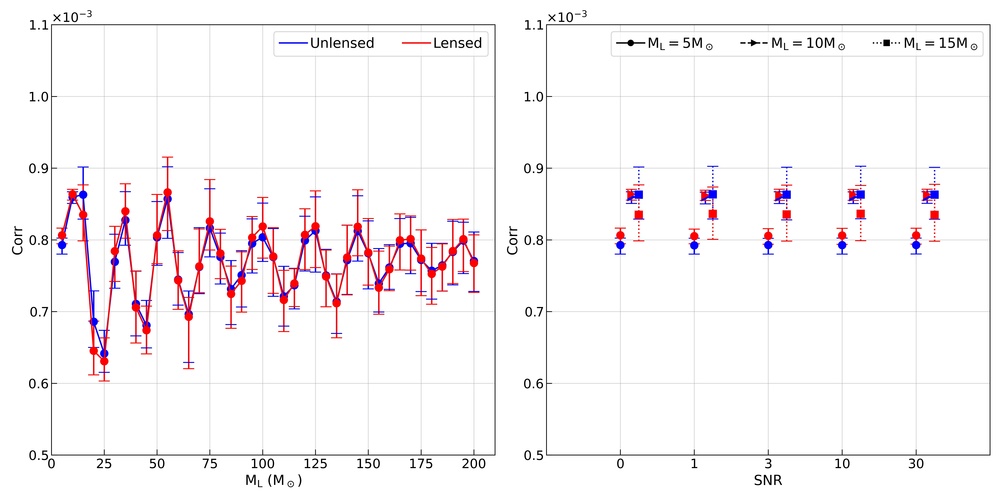}
	\includegraphics[height=5.3cm, width=14cm]{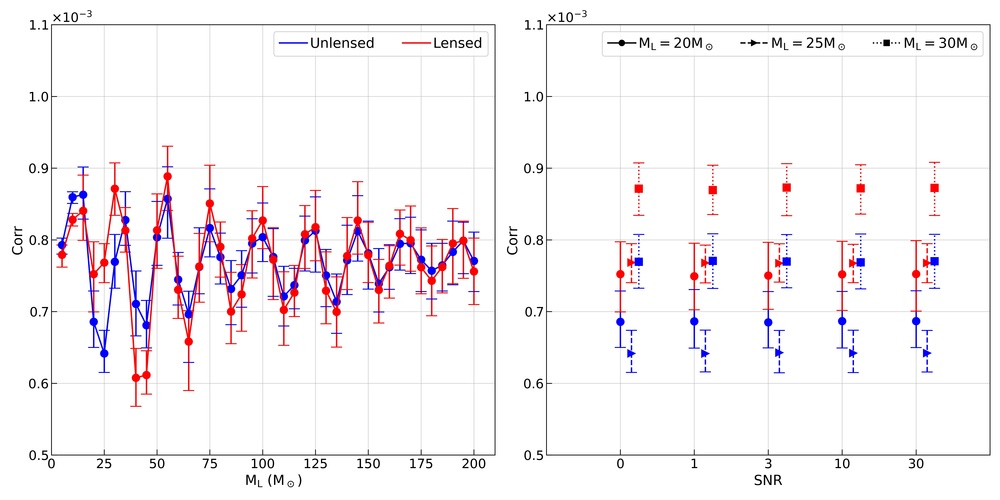}
	\includegraphics[height=5.3cm, width=14cm]{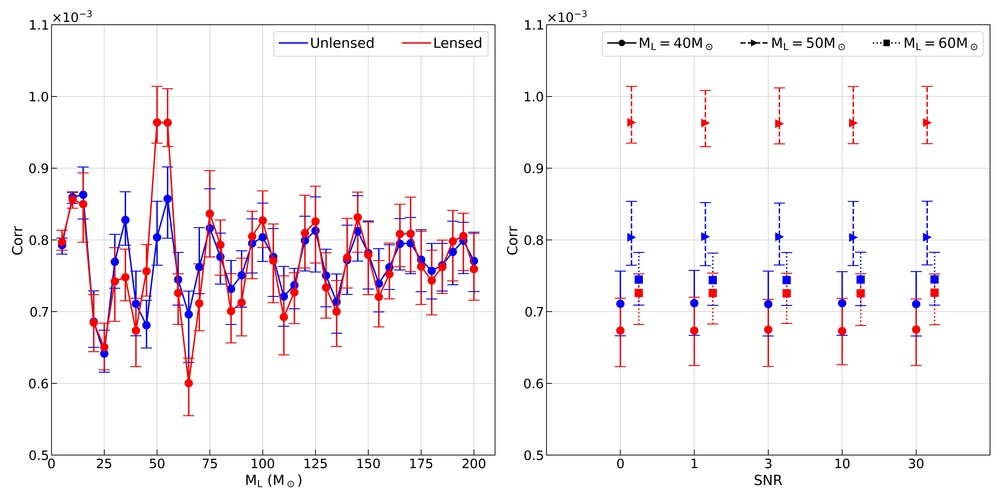}
	\includegraphics[height=5.3cm, width=14cm]{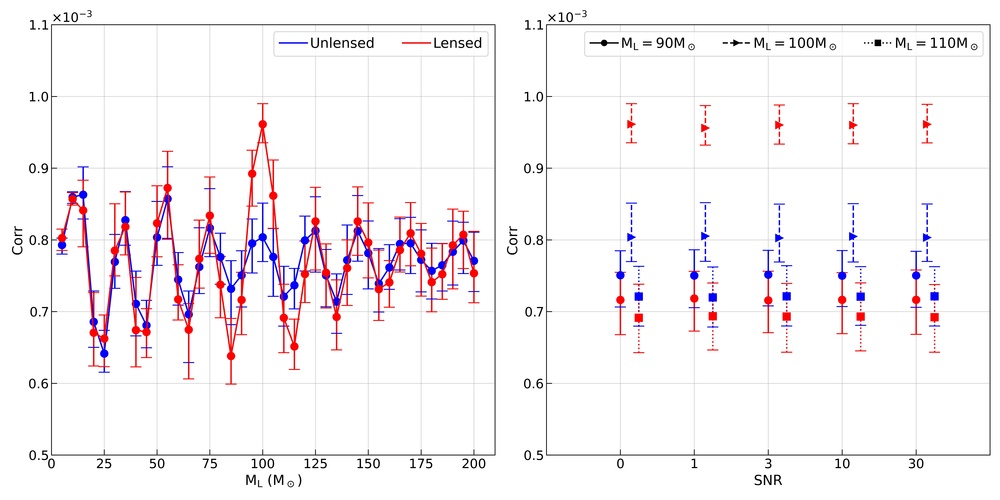}
	\caption{Using the Correlation Integral to estimate M$_{\rm L}$ at 
	known $y(=1)$: Top to bottom, the signals are lensed by a 5, 
	25, 50 and 100 M$_\odot$ lens respectively. The correlation integral 
	is computed by assuming different values of ${\rm M}_{\rm L}$ 
	as indicated on the x-axis. Blue (red) points correspond to 
	unlensed (lensed) signals, solid points are median values, and 
	errors bars correspond to $16^{\rm{th}}$ and $84^{\rm{th}}$ percentiles. 
	Similar to Figures \ref{fig:pk_match_SNR} and \ref{fig:corr_intgl}, the left
	panels correspond to the case without any
	external noise. The right panel represents the effect of different SNR
	values near the actual lens mass value.
	As has been shown in Figure \ref{fig:corr_intgl}, this method 
	is able to differentiate between lensed and unlensed signals only 
	for ${\rm M}_{\rm L} {\geq} 15 {\rm M}_\odot$, and while estimates of 
	${\rm M}_{\rm L}$ can be made in this regime, the estimates are not 
	always accurate unless ${\rm M}_{\rm L} {\geq} 50 {\rm M}_\odot$.}
	\label{fig:corr_intgl_mass}
\end{figure*}

To illustrate some key aspects using a simple example, we consider
white noise as a test signal.
This is broad band and incoherent. 
By definition, such a signal has no frequency structure or evolution
of the power spectrum over time, unlike the expected signal from
CCSN. 
As a result, unlike the Gaussian wave
packet example, it is 
not straightforward to identify differences between the lensed and
unlensed waveforms. The left panel of Figure \ref{fig:white_noise}
shows an example of lensed (red) and unlensed (blue) versions of white
noise. We here take ${\rm M}_{\rm L} = 100 {\rm M}_\odot$. The lensed signal is
shifted vertically for a better comparison between lensed and unlensed
signal. The right panel shows the corresponding power spectra, with
the two curves vertically displaced again for better
visualisation. The modulus of the amplification factor is shown in
black. As can be seen, there is a clear correlation between the power
of a lensed signal and oscillations of the amplification factor. The
unlensed signal, however, shows no such correlation. We use this
correlation as the basis of identifying lensed signals in the
following sections.

\begin{figure*}[ht!]
	\centering
	\includegraphics[height=5.3cm, width=14cm]{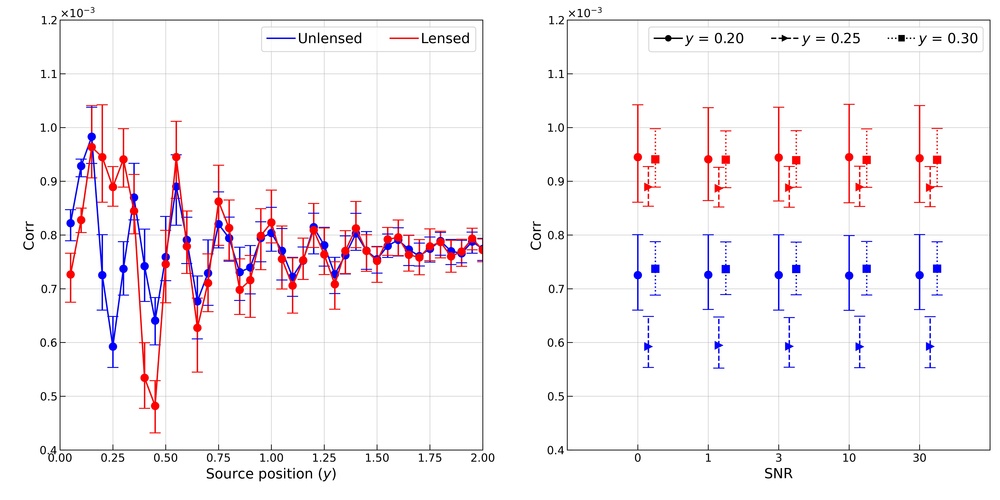}
	\includegraphics[height=5.3cm, width=14cm]{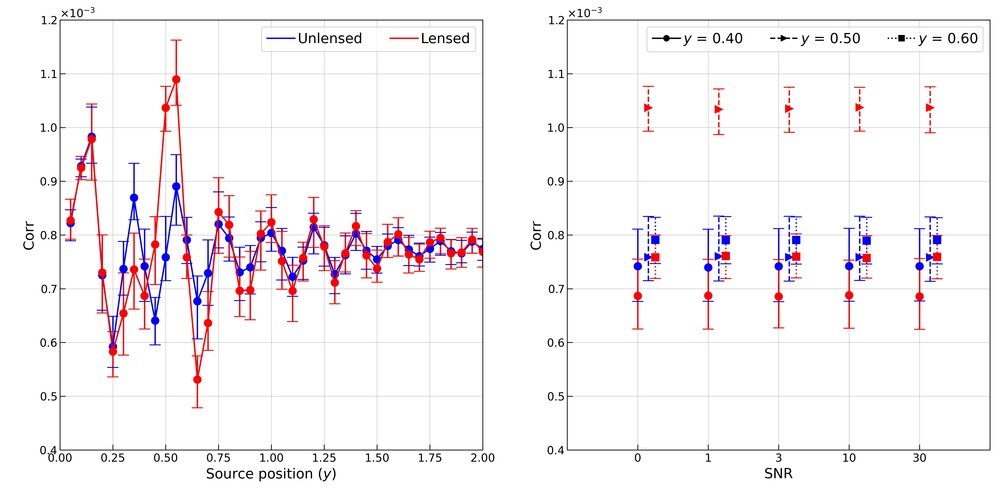}
	\includegraphics[height=5.3cm, width=14cm]{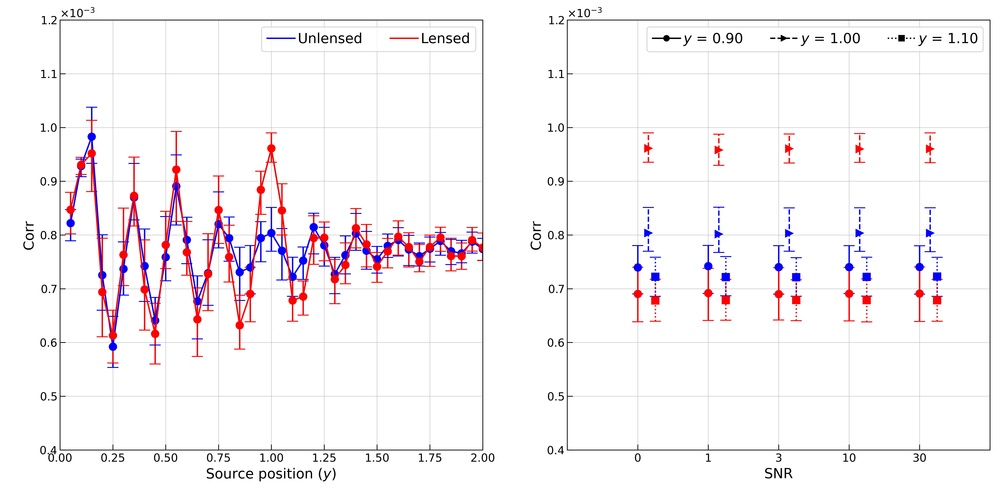}
	\includegraphics[height=5.3cm, width=14cm]{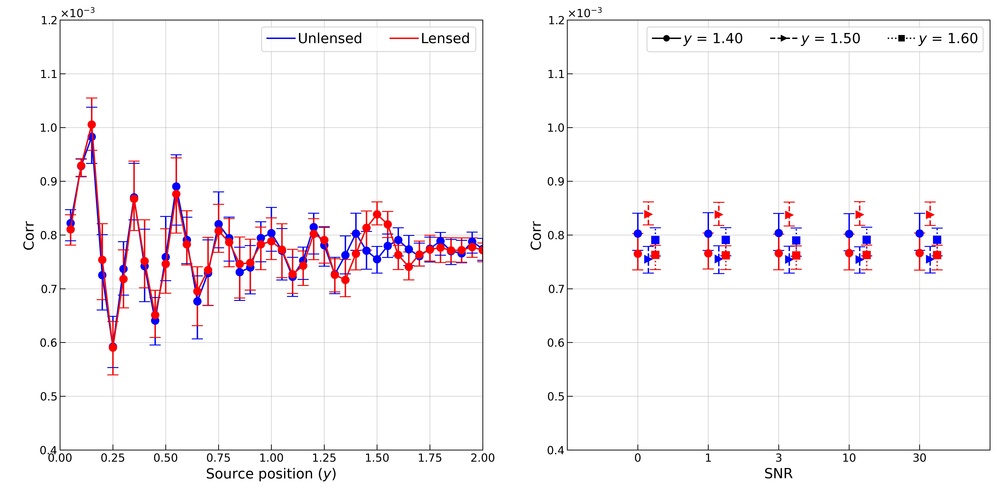}
	\caption{Using the Correlation Integral to estimate $y$ at known 
	${\rm M}_{\rm L}(=100{\rm M}_\odot)$: Top to bottom, the signals 
	are lensed by assuming the source to be present at $y=$ 0.25, 0.5, 
	1.0 and 1.5 respectively. The correlation integral is computed by 
	assuming different values of $y$ as indicated on the x-axis. Blue 
	(red) points correspond to unlensed (lensed) signals, solid points 
	are median values, and errors bars correspond to $16^{\rm{th}}$ and
	$84^{\rm{th}}$ percentiles. The rest of the details are similar to Figure \ref{fig:corr_intgl_mass}.}
	\label{fig:corr_intgl_y}
\end{figure*}

\begin{figure*}[ht!]
	\centering
	\includegraphics[height=6.5cm, width=8.3cm]{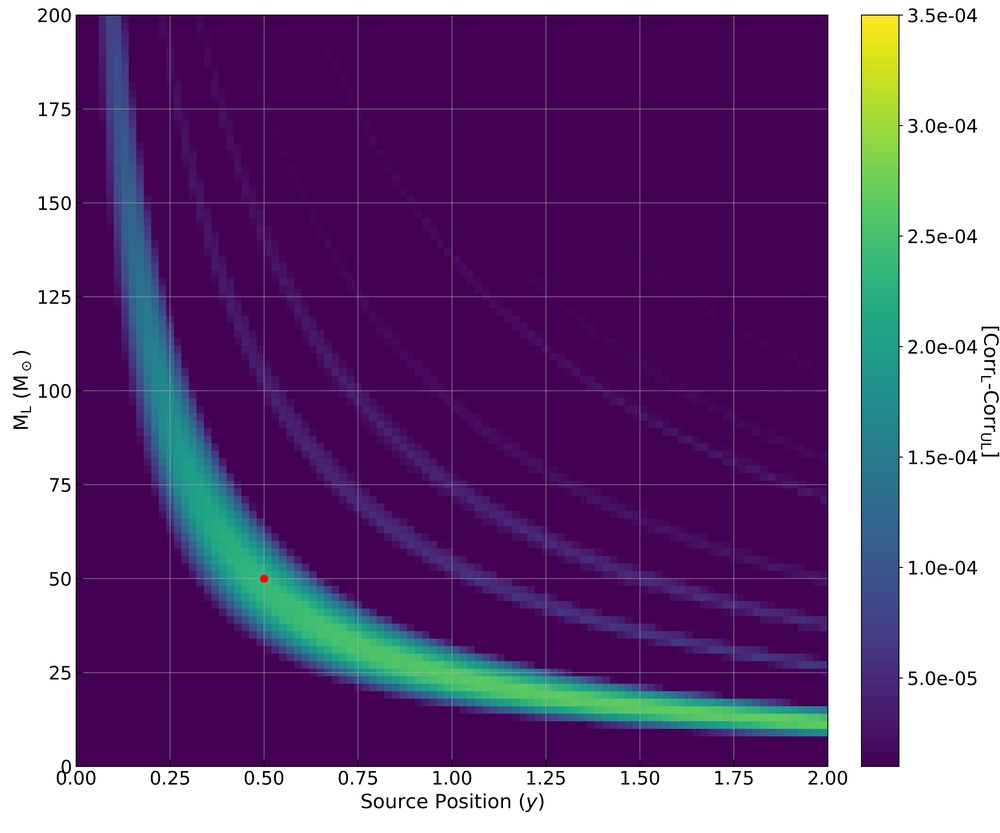}
	\includegraphics[height=6.5cm, width=8cm]{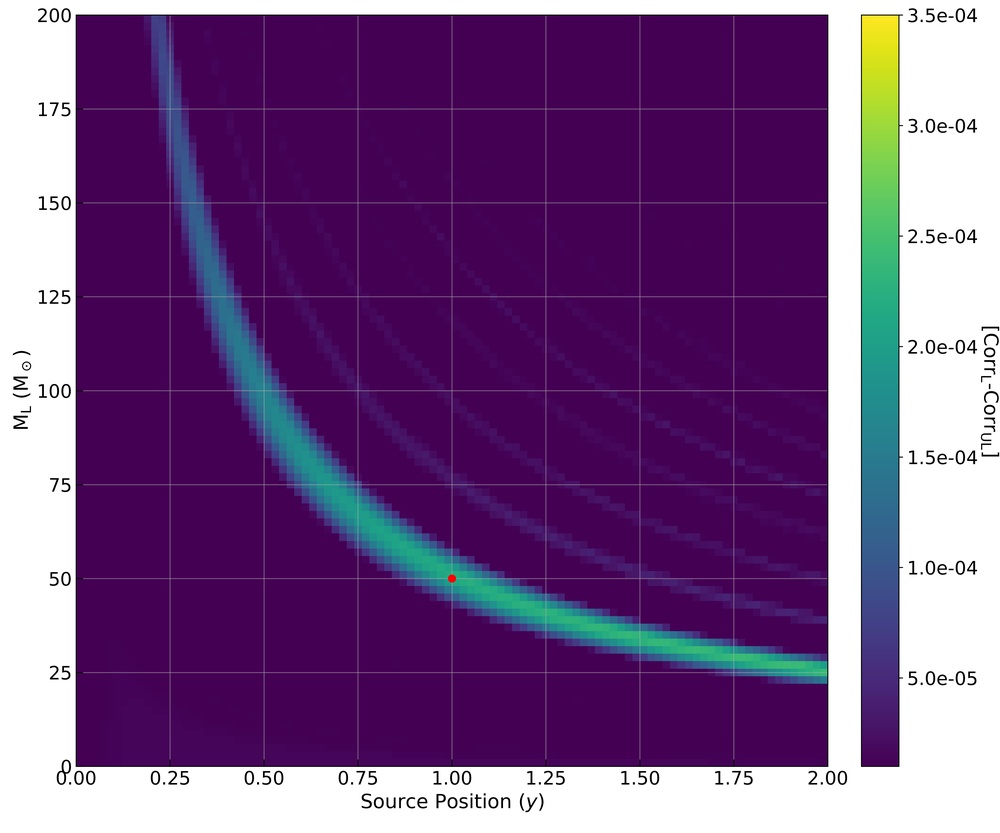}
	\includegraphics[height=6.5cm, width=8.3cm]{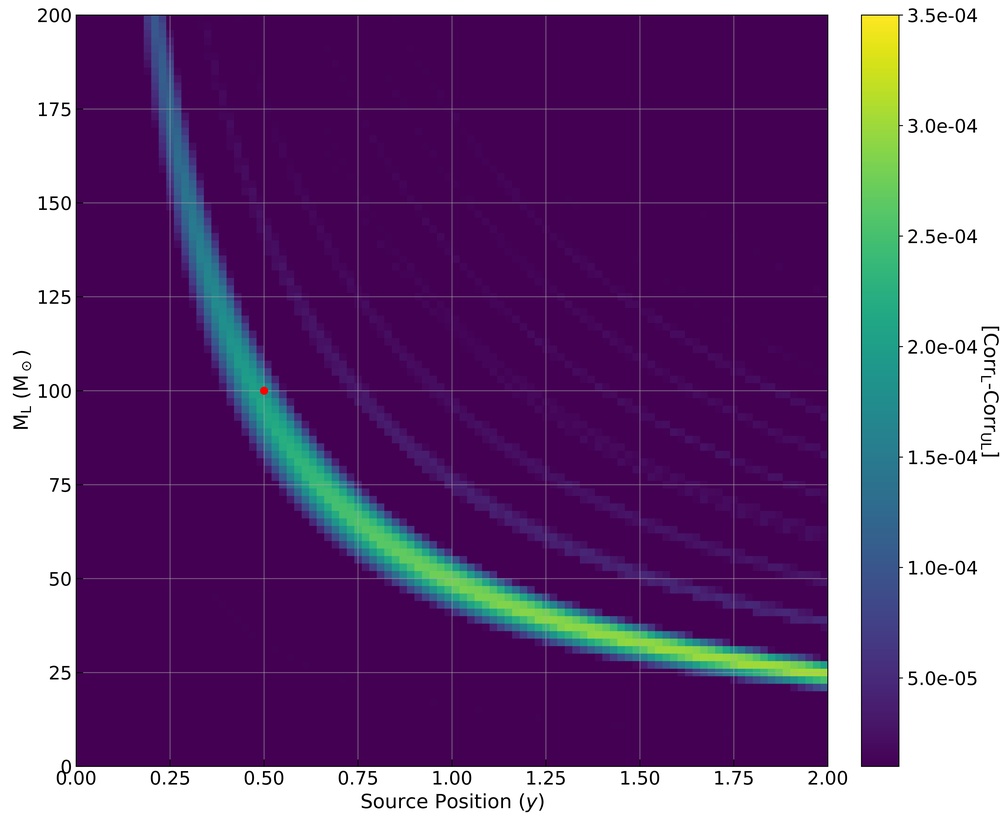}
	\includegraphics[height=6.5cm, width=8cm]{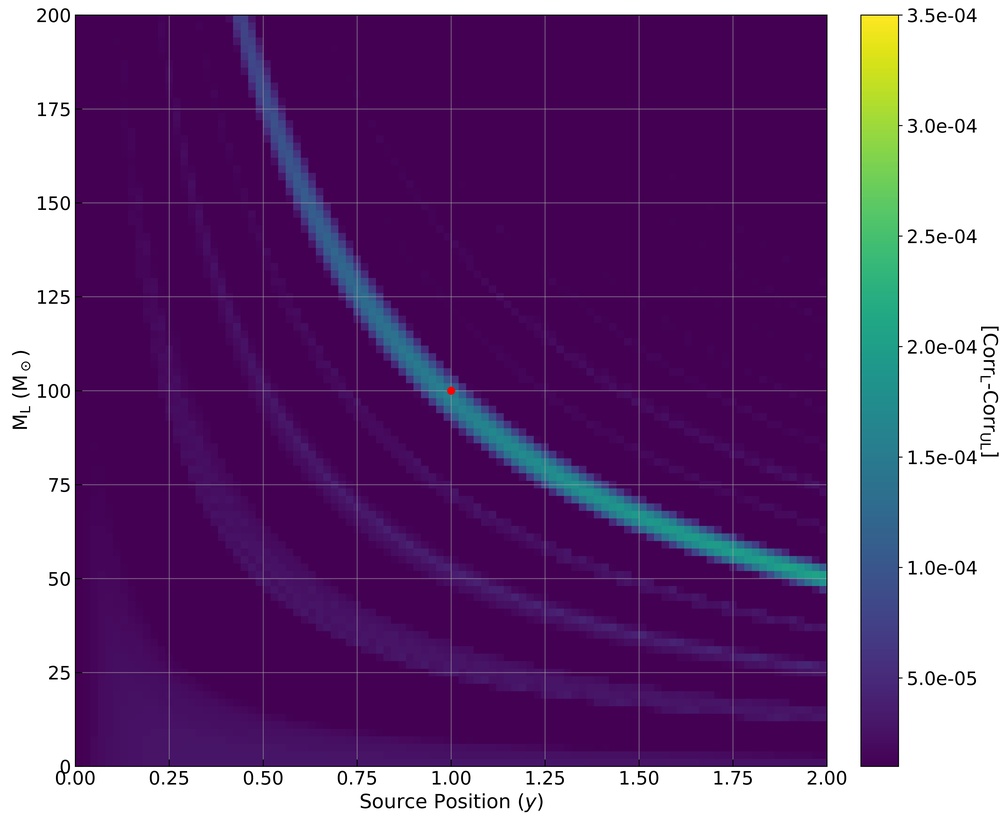}
	\caption{Contour plots for the median value of the correlation 
	integral: In each panel, the reference signal is lensed by the 
	parameter values denoted by the red dot. The correlation integral 
	for the (reference) lensed and unlensed signal is computed by 
	assuming the lens parameters as denoted by the corresponding 
	x- and y-axis values. Each panel shows the difference of
	correlation values between lensed and unlensed signal.}
	\label{fig:corr_intgl_contour}
\end{figure*}

\section{CCSN-like GW Signal}
\label{sec:sig_gen}

Over the last decade, various multidimensional CCSN simulations 
have revealed a multitude of information about different processes
responsible for the emission of photons, neutrinos, and GWs
\citep[e.g.,][]{2008PhRvD..78f4056D, 2008A&A...490..231S, 
2010CQGra..27k4101S, 2013ApJ...766...43M, 2013ApJ...779L..18C,
2015PhRvD..92h4040Y, 2017MNRAS.468.2032A, 2018ApJ...861...10M, 
2019ApJ...876L...9R, 2019MNRAS.486.2238A, 2020MNRAS.494.4665P, 
2020ApJ...898..139W, 2020ApJ...901..108V, 2021MNRAS.503.3552A}.
Unlike chirp signal from merging binaries, the GW signal emitted
in CCSN is broadband in nature covering a frequency range of 
$\sim$(5Hz, 1500Hz).
This broadband signal only lasts around one second.
The exact numbers depend on the mass and spin of the progenitor 
star.
Following the bounce, there are various processes that give
rise to gravitational waves:
\begin{enumerate}
\item
  f-mode and g-mode oscillations of the Proto-Neutron Star (PNS) give
  rise to waves in the frequency range of a few hundreds to thousand
  Hertz. The (relative) intensity of waves produced by this method is
  the highest. 	
\item
  The Standing Accretion Shock Instability (SASI), that is formed as a
  result of the stalling of the shock by the infalling matter,
  accounts for waves in the region of $\sim$100--300Hz. This is
  prominent in only those supernovae events where an explosion does
  not take place.  Prompt convection following the bounce also gives
  rise to waves in the same frequency bracket (for the first $\sim$10
  milliseconds following the bounce). 
\item
  Asymmetric flow of matter and neutrinos produce waves in the regime
  of $\sim$0.1--10Hz. The strain produced here is the weakest amongst
  the various processes highlighted here. 
\end{enumerate}
Owing to the variety of processes involved in the production of GWs, 
these signals may offer a multi-dimensional probe into CCSN events. 
For instance, the ratio of the progenitor star's core kinetic to potential 
energy, post bounce oscillation frequency of the proto-neutron star 
\citep{2021PhRvD.103b3005A}, CCSN explosion mechanism \citep{2016PhRvD..94l3012P}, 
angular momentum of the progenitor \citep{2014PhRvD..90d4001A}, and the  
nuclear equation of state \citep{2021PhRvD.103b4025E} are some of the 
many properties that can potentially be surmised from the observed
gravitational waves.

Despite all the above known information, it is not straightforward to
develop templates for the CCSN GW signals like merging binaries.
In our current work, we use a simple approach (described below) to 
generate CCSN-like mock GW signals.  
We start with a Gaussian Random Field in the frequency domain with 
Power Spectrum $P(f)$:
\begin{equation}
    h(f_j) = (a_j + b_j) \sqrt{P(f_j)},
    \label{eq:freq_domain}
\end{equation}
where, $a_j$ and $b_j$ represent random numbers drawn from a unit 
Gaussian distribution.
To generate the corresponding time-domain signal, we use the 
following relation:
\begin{equation}
    h(t) = \Sigma_{(f_0 - 3 \sigma_{f})}^{(f_0 + 3 \sigma_{f})} h(f) 
    \exp[-2 \pi i f t - (f - f_0)^2/(2\sigma_f^2)]\Delta f,
    \label{eq:time_domain}
\end{equation}
where, $f_0 \equiv f_0(t)$ is the central frequency, and $\sigma_f 
\equiv \sigma_f(t)$ is a measure of the spread of the signal 
(in frequency domain) around the central frequency.
This time dependent filtering scheme removes contribution of
frequencies far away from $f_0(t)$ where the frequency separation is
quantified in units of $\sigma(t)$. 
These functions can be used to model variation of the signal in
frequency space as a function of time.  
Following Equation \ref{eq:time_domain}, we choose values 
such that the central frequency ($f_0$) of the signal increases 
linearly from $100$~Hz to $1000$~Hz in a span of $0.4$ seconds. 
After that, the signal stabilizes at $1000$~Hz.
The spread of signal ($\sigma_f$) increases linearly from $25$~Hz to 
$100$~Hz during the first half of the signal, and drops linearly from 
$100$~Hz to $25$~Hz during the second half.

Figure \ref{fig:ccsn_gw} represents one example of a CCSN-like 
signal generated using the above described method. 
The left panel shows the signal in time domain with time on
x-axis and signal strain on y-axis. 
The right panel represents the corresponding (time-frequency)
spectrogram. One can notice the linear increase between time
and frequency and the saturation above $0.4$~seconds.

\section{Results}
\label{sec:results}

In this section, we present our results of analysis of gravitational
lensing of CCSN-like mock GW signals generated using the method
described in \S\ref{sec:sig_gen}.

\subsection{Lensed CCSN Signal}
\label{ssec:ccsn_lensed}

The current and upcoming GW detectors cannot detect the cosmological
CCSN GW signals. 
Only signals coming from within our own Galaxy or the 
nearby satellite galaxies are likely to be detected. 
Hence it is the stellar mass compact objects that can introduce the
lensing effects in these CCSN GW signals which can be very well
approximated by a point mass lens (see \ref{sec:appendix_A} for
further details).
In principle one can consider intermediate mass black holes but the 
number of such objects is expected to be very small.

In Figure \ref{fig:ccsn_gw_ps}, we present two lensed CCSN-like GW
signals.
The top left (right) panel represents the lensed and unlensed 
signals in red and blue, respectively, with a lens of 20M$_{\odot}$
(100M$_{\odot}$).
As has been discussed in \S\ref{sec:gaussian}, one can see that the 
lensing introduces both amplification and
de-amplification based on the frequency value of the signal.
Lensing also shifts the signal in time (phase shift) although
not visible in this plot but it can be inferred from Figure 
\ref{fig:point_lens} or Figure \ref{fig:gaussian}.
The bottom left (right) panels represent the power spectrum of 
lensed and unlensed with lens mass 20M$_{\odot}$ (100M$_{\odot}$).
The solid black line represents the amplification factor ($|F|$)
due to the lens mass.
One can clearly see a correlation between the
amplification factor ($|F|$) and the amplitude of peaks in the
corresponding (lensed) power spectrum: in the power spectra of 
lensed signals, peaks at frequencies surrounding maximas of the 
amplification factor are amplified,  and peaks at frequencies 
surrounding minimas of the amplification factor are de-amplified. 
No such trend is observed for power spectra of unlensed signals.

\subsection{Identifying a Lensed Signal}
\label{ssec:lensed_identify}

\subsubsection{Tracking Oscillations in Power Spectra}
\label{ssec:track_osc}

While the power spectrum seems to be a promising tool to 
track down a lensing event, one would like to know how 
efficient this method is, or whether any statistical 
statements can be made about using such a method. 
One way to address this question is to estimate the 
average distance between a peak of the amplification factor 
and the closest peak of the power spectrum. 
One can alternatively focus on the correlation between troughs
in the amplification factor and the observed signal. 
Both these methods yield qualitatively similar results.
We first discuss an approach based on the peaks alone. 
There is one main problem with the proposed method: owing to 
the stochastic nature of the signal, there are a large number 
of `false peaks', i.e. peaks arising as a result of the 
randomness associated with the signal (in our case, this arises due to 
the white noise power spectrum and the Gaussian random field), 
in both the lensed and unlensed power spectra, 
which would naturally bias our results. 
If external noise is present, this adds to the number of 
apparent peaks, which may further bias our results. 

To tackle the above-mentioned issue of false-peaks, we employ the
following method to estimate  
`genuine peaks': we first use a Savitzky-Golay filter 
\citep[][hereafter \textsc{savgol}]{1964AnaCh..36.1627S} with 
a bin size of $N = 11$ to smoothen out the power spectrum. Although
arbitrary, the value of $N$ is important: if $N$ is too small, the
smoothened curve retains a large amount of randomness of the signal, while
if $N$ is too large, the number (location) of peaks drastically
reduces (shifts). After experimenting with a few different values, we
find that $N = 11$ works well, and we hence proceed with this. 
For the highest peak of the smoothened power spectrum, we 
traverse to higher and lower frequencies such that the power 
is half of the power at the peak. 
The location of the peak is then estimated as a weighted average 
across the bounded region obtained above: 
$f_{pk} = \frac{\Sigma P(f) f}{\Sigma P(f)}$. 
We repeat this process for all other peaks, selecting peaks 
in descending order.
Figure \ref{fig:ccsn_gw_ps_smooth} shows the results of the above peak 
finding algorithm for the power spectrum of the unlensed signal from
Figure \ref{fig:ccsn_gw}.  
The top panel in Figure \ref{fig:ccsn_gw_ps_smooth} represents
the power spectrum of the unlensed signal in blue and the
corresponding smoothed power spectrum in black.
The middle (lower) panel represents the lensed and smoothed
power spectrum for the case with a lens mass of 20M$_\odot$ 
(100M$_\odot$).
The black `$\times$' symbols represent the frequency values
corresponding to the peaks estimated using the above mentioned algorithm.
Relative to estimated peaks of the unlensed signal, one can see that
the lensing due to a 20M$_\odot$ lens increases the number of peaks in
the ${<}600$Hz range, whereas the number of peaks are reduced in
${>}600$Hz region. 
The same is true for lensing due to a 100M$_\odot$ lens, although
in different frequency regions due to rapid oscillations of
the corresponding amplification factor.
Overall, the number of peaks of the original power spectrum is reduced
considerably, which helps counter the bias of randomness of the signal, at
least partially. Throughout the rest of the paper, we continue to
refer to peaks identified using this algorithm as `genuine peaks'. 

We now proceed to check the efficiency of the previously described
power spectrum method to identify lensing: we generate 100 mock
signals using the method outlined in \S\ref{sec:sig_gen}, all with the
same power spectrum.
For each signal, we estimate the average distance between peaks of the
amplification factor and estimated `genuine-peaks' of the power
spectrum, for both lensed and unlensed signals.
This quantity is denoted by $\Delta f$ in Figure
\ref{fig:pk_match_SNR}. Blue (red) points correspond to the results of
the unlensed (lensed) signals, the solid points are median values, and
the error bars correspond to $16^{\rm{th}}$ and $84^{\rm{th}}$
percentiles.
We consider multiple cases with varying amounts of external white noise 
injected into the mock signal. 
Since our method is not optimized to identify instrumental noise, we include 
time-independent white noise representing instrumental noise only in
the same frequency range where the  
original signal is present, 
i.e. in the frequency range ${\sim} [100{\rm Hz}, 1100{\rm Hz}]$.  
This ensures that our results are not biased by those regions which
only contain external noise.
We begin by analyzing the case with no external noise (left panel of
Figure \ref{fig:pk_match_SNR}), and later extend the discussion to the
other cases (right panel of Figure \ref{fig:pk_match_SNR}). 
As one can see from the left panel, for relatively less massive lenses, 
the median values are well within the two percentile regions. 
As the mass of the lens is increased, the two median values separate 
out from the percentile regions (at ${\sim} 100 {\rm M}_\odot$), and 
ultimately for higher masses, the two percentile regions decouple as 
well (at ${\sim} 150 {\rm M}_\odot$).
This shows that although the outlined power spectrum tool is useful, 
it is dependable only for high mass lenses (${>} 150 {\rm M}_\odot$). 
This corresponds to cases where the amplification factor
undergoes multiple oscillations within the frequency-range of the
signal, and hence a greater number of (more rapid) oscillations are
observed in the power spectrum of the (lensed) signal. It is for the
same reason that the error bars begin to diminish as ${\rm M}_{\rm L}$ increases:
for heavier lenses, since there are a greater number of maximas in the
considered frequency range, the number of `distances' under
consideration is larger. Computing the average of a greater number of
`distances' would smoothen out the variation, resulting in tighter
error bars. While this argument holds for both lensed and unlensed
signals, there is one difference: we observe a greater number of
`true' peaks for the lensed signal, while we observe `random' peaks
for the unlensed signal. That is why we observe a steady decrease with the
median values of the lensed signal, while the median values of the
unlensed signal oscillate randomly.  
As noted earlier, this stochastic behavior (of the unlensed signals)
stems from the randomness associated with the original signal.

We next proceed to see whether our power spectrum method is able to 
identify lensed signals when external noise is present in the signal
as well (right panel of Figure \ref{fig:pk_match_SNR}). 
Although this plot demonstrates trends for only three different 
values of $\rm{M}_{\rm{L}}$, the following discussion takes all other 
masses (shown in the left panel) into account as well. 
For SNR = 30, the results are identical to those corresponding to 
case with no added noise. 
For SNR = 10, we note differences for the lensed results only at small
values of $\rm{M}_{\rm{L}}$. 
This is expected, as these are the regions where trends carry the 
imprint of the inherent randomness of the signal. 
However, the results of the unlensed signal show variations at all 
values of $\rm{M}_{\rm{L}}$. 
For smaller values of SNR, the median values show some variation for 
both lensed and unlensed signals at all lens masses. 
Despite the variation, at large $\rm{M}_{\rm{L}}$, the median lensed
values are all comparable to each other.  
With the percentile regions, we note an important difference between 
low and high SNR scenarios: for high SNR (10 and 30), the two 
percentile regions begin to separate out at $150 {\rm M}_\odot$, which is 
consistent with the case with no external noise. 
However, this separation is observed at $155 {\rm M}_\odot$ ($160 {\rm
  M}_\odot$) for the case with SNR = 3 (SNR = 1). 
We attribute this to the excess noise that pollutes the features of 
the power spectrum.

\subsubsection{Correlation Integral}
\label{ssec:corr_int}

While studying the average distance between the peaks of the power
spectrum and amplification factor is one way to proceed, one may also
look at the correlation integral between the amplification factor and
the power spectrum. Specifically, the following integral can be
studied: 
\begin{equation}
   {\rm Corr} :=  \frac{\int df\ P(f) |F|^2(f)}{\left[\int df\ P(f)
       \right] \left[ \int df\ |F(f)|^2\right]},
   \label{eq:corr_int}
\end{equation}
where $P(f)$ is the  power spectrum and $|F(f)|$ is the absolute 
value of the amplification factor at a given frequency value $f$.
The advantage is that we do not have to use any techniques to estimate
genuine peaks of the power spectrum.
The left panel of Figure \ref{fig:corr_intgl} shows the above integral 
estimated at different values of ${\rm M}_{\rm L}$ for the case with 
no external noise.
As in Figure \ref{fig:pk_match_SNR}, blue (red) points correspond
to the results of the unlensed (lensed) signals, the solid points are
median values, and the error bars correspond to $16^{\rm{th}}$ and
$84^{\rm{th}}$ percentiles.
As can be seen, except
for ${\rm M}_{\rm L} {\leq} 10 {\rm M}_\odot$, there is a clear 
offset between the median values of the lensed and unlensed signals, 
and the percentile regions are well separated as well.
While an oscillatory
behavior is expected for the median values of the unlensed signals,
one expects the median values of the lensed signals to increase
monotonically with lens mass, since the power spectrum displays more
rapid oscillations at higher values of ${\rm M}_{\rm L}$.
Instead, we observe that
the correlation integral for the lensed signals rise to a point, and
more or less flatten out. We attribute this behaviour to the randomness of the CCSN-like GW signal. 
In the right panel, we consider multiple cases with varying values of 
SNR, and follow the same procedure to inject time-independent white
noise as outlined in the previous Subsection \ref{ssec:track_osc}. 
Once again, we show results only for three values of ${\rm M}_{\rm L}$, 
but the following statement holds true for all other cases as well: 
the results remain more or less the same for a wide range of values 
of SNR, and the variation between cases with different SNR is smaller 
than the variation observed in the previous peak-matching method.  

\subsection{Constraining Point Mass Lens Parameters}
\label{ssec:constraint}

While the two methods outlined above offer possible avenues to 
identify lensed signals, they require a priori knowledge of the 
behaviour of the amplification factor. 
For an isolated point mass lens, this corresponds to knowing the 
values of the lens mass ($\rm{M}_{\rm{L}}$) and the source position ($y$). 
While this stands true for the peak matching method, the correlation 
integral method is able to identify lensed signals even in cases 
where an estimate of $\rm{M}_{\rm{L}}$ is not available, as long as 
$y$ is known. 
To demonstrate this, in Figure \ref{fig:corr_intgl_mass}, from top 
to bottom, we consider signals to be lensed by a 5, 25, 50 and 100 
${\rm M}_\odot$ lens respectively. 
In all cases, we fix $y=1$. The correlation integral is estimated at 
multiple values of $\rm{M}_{\rm{L}}$, as indicated on the x-axis. 
Consistent with our observations from Figure \ref{fig:corr_intgl}, 
this method is not able to identify lensing when $\rm{M}_{\rm{L}} 
{\leq 10} {\rm M}_\odot$. 
From the bottom two panels, we notice that the correlation integral 
peaks near the true value of $\rm{M}_{\rm{L}}$. 
It is also only around the true value of $\rm{M}_{\rm{L}}$ that 
(a) there is a significant difference between the correlation of 
lensed and unlensed signals, and 
(b) the percentile regions of the lensed and unlensed signals separate out. 
We also note that there is a relatively sharp drop in the correlation 
on either side of the peak.
In addition to identifying lensed signals, such an approach may be 
useful to obtain an estimate of $\rm{M}_{\rm{L}}$.

In the second panel from the top, although the correlation integral 
does peak around the true value of $\rm{M}_{\rm{L}}$, the peak is not 
a global maxima. 
As can be seen from the plot, the reason for the same is that the 
(median) correlation integral of the unlensed signals corresponding to
the true value of $\rm{M}_{\rm{L}}$ (=$25{\rm M}_\odot$) is a
minima. It thus seems that the accuracy of the estimate of
$\rm{M}_{\rm{L}}$ depends on the value of the correlation integral of
the unlensed signals, at least for small $M_L$. From Figure
\ref{fig:corr_intgl}, for $\rm{M}_{\rm{L}} {\geq} 50 {\rm M}_\odot$,
we see that the (median) correlation integral of all lensed signals is
always greater than the \textit{maximum} (median) correlation integral
of unlensed signals. For $\rm{M}_{\rm{L}} {\geq} 50 {\rm M}_\odot$, a
global maxima will hence be present when one tries to make a plot
similar to Figure \ref{fig:corr_intgl_mass}. Thus, while it may be
possible to estimate $\rm{M}_{\rm{L}}$ for $\rm{M}_{\rm{L}} {>} 10
{\rm M}_\odot$, the estimate may not always be accurate unless
$\rm{M}_{\rm{L}} {\geq} 50 {\rm M}_\odot$. Especially for $10 {\rm
  M}_\odot {<} \rm{M}_{\rm{L}} {<} 50 {\rm M}_\odot$, in addition to
locating maximas, it may be useful to check if there is a sharp drop
in the correlation on either side of the peak. This may provide
additional hints as to whether one is close to the true value of
$\rm{M}_{\rm{L}}$. 
At this juncture, we find it important to state the following point:
the above mentioned `limits' of $10 {\rm M}_\odot$ and $50 {\rm
  M}_\odot$ were obtained with $y$ fixed at 1. Since the amplification
factor oscillates slower (faster) for smaller (larger) values of $y$,
the above mentioned limits would be larger (smaller) for smaller
(larger) values of $y$. Note that the value of $y$ will similarly
affect the `limits' of the peak-matching method as well.

Another case that one can consider is when an estimate of 
$\rm{M}_{\rm{L}}$ is available, but $y$ is unknown. 
Figure \ref{fig:corr_intgl_y} explores this scenario. 
From top to bottom, we consider the source to be positioned at 
$y=$ 0.25, 0.5, 1.0 and 1.5 respectively. 
In all cases, we fix $M_L=100{\rm M}_\odot$. 
The correlation integral is estimated at multiple values of $y$, 
as indicated on the x-axis. 
Although the percentile regions of the lensed and unlensed signals 
separate close to the actual value of y, other regions also show 
significant separation between lensed and unlensed signals
(for example $y$ = 0.25, 0.50 cases). 
Such an observation indicates the possible degeneracies in the 
lens parameter values.

The final case that we explore is a rather extreme one in which 
both $\rm{M}_{\rm{L}}$ and $y$ are unknown. 
In Figure \ref{fig:corr_intgl_contour}, we see if the correlation 
integral method can be used for performing estimation of these two 
parameter simultaneously. 
In each panel, the reference signal is lensed using parameters 
denoted by the red dot. 
The correlation integral is computed by assuming various lens 
parameters as denoted by the corresponding x- and y-axis values. 
Figure \ref{fig:corr_intgl_contour} shows the median value of the 
correlation integral. 
Note that no external noise is injected into any of these signals. 
In each panel, the value of difference of correlation
integral between lensed and unlensed signal is plotted, i.e.,
$\rm{[Corr_{Lensed} - Corr_{Unlensed}]}$. 
Figure \ref{fig:corr_intgl_contour} is a two-dimensional analog
of Figure \ref{fig:corr_intgl_mass} and \ref{fig:corr_intgl_y}.
While the true set of parameters is always close to contours
with high values of the correlation integral, the contours are not 
tight enough to provide a reasonable estimate of the two parameters. 
This is due to the degeneracies between the two parameters.
Such a result is also expected from Figure \ref{fig:point_lens},
where we have shown the amplification curves while fixing one of these
parameters. 
From the top-left panel, one can see that if both $y$ and $\rm{M_L}$
have small values, then it can be tough to end up near the actual 
values as the lensing does not introduce significant changes in 
the signal.
Even if we do increase the mass of the lens, the banana shape of
the contours allows a wide range of the lens parameters.
Hence, we do not expect to tightly constrain both parameters
simultaneously in most cases.

\subsection{Limitations of our Methods}
\label{ssec:limitations}

We have considered two methods to establish presence of gravitational
lensing.
The proposed peak-matching method is reliable only for massive lenses
(M$_{\rm L} {>} 150{\rm M}_{\odot}$). 
The correlation integral method is able to differentiate between 
lensed and unlensed signals starting at M$_{\rm L} {\sim} 
15{\rm M}_{\odot}$, but even such lenses are unfortunately not very
abundant. 
For a supernova within the galactic plane, juxtaposition of multiple
lenses may lead to a high effective value for the lens mass.
Apart from this limitation, we summarise a few other
limitations of our methods.

In the previous subsection, we have introduced a possible method to
estimate genuine peaks. Peaks identified using this method seem to
provide results that are well aligned with what one would intuitively
expect, at least for massive lenses. However, we note that this
proposed method is not entirely foolproof: when one chooses to create
a bounded region around each peak such that the power within the
region is within $50\%$ of the value of the peak, the number of
estimated peaks remains more or less constant across all
signals. As the mass of the lens increases, the number of peaks
corresponding to the amplification factor increase, and at some
point, the number of peaks of the amplification factor equals the
number of estimated peaks of the power spectrum. For lens masses above
this limit, the value of $\Delta f$ will worsen due to the
limited number of peaks. This is the reason why the upper limit of the
error bar widens for the last few (lensed) data points in Figure
\ref{fig:pk_match_SNR}. In such a case, we  have to explore whether
changing the smoothing scale (number of bins) or the level up to which
we average (half power in the results presented here) the signal has an impact. 

Another way to reduce the number of false peaks is to consider only
those peaks of the smoothed power spectrum that are larger than a
given lower threshold (for e.g., three times the rms value). However,
introducing such a threshold seems to greatly reduce the number of
peaks estimated using our method, and the values of $\Delta f$ are
biased as a result.
In our discussion, we have only considered white noise as external noise. 
By definition, white noise has almost equal power at all frequencies. 
In real life observations, noise follows a different power
spectrum as different sources of noise dominate for different 
frequencies.
Modelling with frequency dependent noise is required to study if it
introduces any non-trivial effects.
However, as the instrument noise does not have any oscillatory
behaviour in frequency, we do not expect any significant changes. 

It is noteworthy that the error bars in Figure \ref{fig:pk_match_SNR}
are asymmetric with respect to the median value. This arises due to
the limited resolution of the power spectrum: the resolution here is
$\sim$2Hz, because of which there is a bias with respect to the number
of `small values' of distances between maximas of the amplification
factor and peaks of the power spectrum that are observed. This leads
to error bars that are shorter on the lower end, which are
noticeable when the median value of $\Delta f$ is relatively
small. Zero-padding the signal before computing the power spectrum
helps in improving frequency resolution, but this dilutes the
number/location of peaks. In addition, during real-life observations,
where sampling frequency of the detector in time domain is limited,
features of the power spectrum may not always be well reproduced. 

An alternate approach to identify lensing is the correlation integral. 
As already noted, the advantage here is that one does not have to 
worry about identifying genuine peaks. 
This method seems to work well as long as the frequency range 
analyzed contains both the `actual' signal and external noise. 
If the analyzed frequency range contains a sub-range in which only 
external noise is present, the correlation will reduce. 
While performing such an analysis, it is thus important to identify 
the frequency range in which the `actual' signal is present, and compute 
the correlation only in this `isolated' region. 
For example, the current LIGO band covers the frequency range of a 
few tens of Hertz to a few thousand Hertz, and when a signal is 
detected, the observed signal (`actual' signal + external noise) would 
cover a majority of this band. 
However, if the actual signal is present only in the frequency range of 
${\sim}[100{\rm Hz}, 1100{\rm Hz}]$, the correlation is to be computed only 
in this range.
Unfortunately, this can be reliably performed only when the SNR is high. 
Such isolation is required for the peak matching method as well.
Further studies are required to explore limitations arising from this
aspect. 

\section{Conclusions}
\label{sec:conclusions}

We have discussed gravitational lensing of wavepackets and broad band
gravitational wave signals in this paper.
We have shown the expected effects of micro-lensing on the amplitude
and phases.
We have demonstrated that the power spectrum of the observed signal is
a useful quantity in the analysis of signal from sources such as core
collapse supernovae. 

We have described two possible methods to identify lensing in
cases where our knowledge of the waveform and power spectrum is
limited.  
Specifically, we have studied cases where the wave nature
of radiation is important during gravitational lensing.
Lensing in
this regime is chromatic, and different frequency components are
amplified/de-amplified by varied factors, as governed by the
oscillations of the amplification factor.
This gives rise to
oscillations in the power spectrum of a lensed signal.
Our proposed methods rely on studying such oscillations.
When the waveform under consideration is broadband and incoherent, as
is the case with a CCSN GW signal, comparison of peaks between power
spectra and the amplification factor is unfortunately dependable only
for massive lenses.
This is due to the presence of peaks in the power spectrum due to
randomness. 
To partially overcome this problem, we employed a possible method to
estimate `genuine peaks'.
This certainly offers an improvement over directly selecting peaks
from the power spectrum. 
However, as has already been noted, this method is not always robust,
and alternate algorithms are required for better analysis.

We have demonstrated that the correlation integral is a powerful tool
to identify gravitational lensing.
This is capable of doing so reliably for lens masses as low as
$15$~M$_\odot$.
Most importantly, this is insensitive to the presence of noise.
We have shown that in principle we can infer parameters of the lens
using the correlation integral though this is not always reliable if
both the lens mass and displacement $y$ are unknown. In the GAIA era, it may 
be possible to get at least partial information about the lens if the 
source is detected using electromagnetic waves as well and is localized 
on the sky.

The rate of SNe in the Large Magellanic Cloud (LMC), the Small Magellanic 
Cloud (SMC) and within the Galaxy is ${\sim}1/200$ year 
\citep{2017ApJS..230....2B}, ${\sim}1/500$ year \citep{vink2020physics} and 
${\sim}2/100$ year \citep{2013ApJ...778..164A}, respectively. 
In addition, the microlensing optical depth for LMC/SMC is ${\sim}10^{-7}$
\citep[e.g.,][]{2005ApJ...633..906B, 2010GReGr..42.2047M, 2011MNRAS.416.2949W}, 
whereas the microlensing optical depth within the Galaxy peaks in the 
Galactic plane (${\sim}10^{-6}$) and continuously decreases at high latitudes
\citep[e.g.,][]{2019ApJS..244...29M, 2020ApJS..249...16M}.
Hence, the chances of observing a microlensed SNe are similar within
the Galaxy and the Magellanic Clouds, although all cases lead to very
small optical depths for microlensed SNe.

The power spectrum method can be applied to any broad band signal and
is especially useful for incoherent signals where the phase
information is not available from models.  

\section*{Acknowledgements}

RR would like to thank Ambresh Shivaji for helpful discussions 
and comments. 
RR also thanks the Department of Science and Technology, 
Government of India for being awarded the INSPIRE scholarship. 
AKM would like to thank Council of Scientific and Industrial 
Research (CSIR) India for financial support through research 
fellowship No. 524007.
Authors thank the anonymous referee for insightful comments.
Authors acknowledge the use of IISER Mohali HPC facility. 
This research has made use of NASA's Astrophysics Data System 
Bibliographic Services.


\bibliography{references}

\begin{thebibliography}{}
\expandafter\ifx\csname natexlab\endcsname\relax\def\natexlab#1{#1}\fi

\bibitem[{{Abbott} {$et~al$.}(2019){Abbott}, {Abbott}, {Abbott}, {Abraham},
  {Acernese}, {Ackley}, {Adams}, {Adhikari}, {Adya}, {Affeldt}, {Agathos},
  {Agatsuma}, {Aggarwal}, {Aguiar}, {Aiello}, {Ain}, {Ajith}, {Allen},
  {Allocca}, {Aloy}, {Altin}, {Amato}, {Ananyeva}, {Anderson}, {Anderson},
  {Angelova}, {Antier}, {Appert}, {Arai}, {Araya}, {Areeda}, {Ar{\`e}ne},
  {Arnaud}, {Arun}, {Ascenzi}, {Ashton}, {Aston}, {Astone}, {Aubin}, {Aufmuth},
  {AultONeal}, {Austin}, {Avendano}, {Avila-Alvarez}, {Babak}, {Bacon},
  {Badaracco}, {Bader}, {Bae}, {Baker}, {Baldaccini}, {Ballardin}, {Ballmer},
  {Banagiri}, {Barayoga}, {Barclay}, {Barish}, {Barker}, {Barkett}, {Barnum},
  {Barone}, {Barr}, {Barsotti}, {Barsuglia}, {Barta}, {Bartlett}, {Bartos},
  {Bassiri}, {Basti}, {Bawaj}, {Bayley}, {Bazzan}, {B{\'e}csy}, {Bejger},
  {Belahcene}, {Bell}, {Beniwal}, {Berger}, {Bergmann}, {Bernuzzi}, {Bero},
  {Berry}, {Bersanetti}, {Bertolini}, {Betzwieser}, {Bhandare}, {Bidler},
  {Bilenko}, {Bilgili}, {Billingsley}, {Birch}, {Birney}, {Birnholtz},
  {Biscans}, {Biscoveanu}, {Bisht}, {Bitossi}, {Bizouard}, {Blackburn},
  {Blackman}, {Blair}, {Blair}, {Blair}, {Bloemen}, {Bode}, {Boer}, {Boetzel},
  {Bogaert}, {Bondu}, {Bonilla}, {Bonnand}, {Booker}, {Boom}, {Booth}, {Bork},
  {Boschi}, {Bose}, {Bossie}, {Bossilkov}, {Bosveld}, {Bouffanais}, {Bozzi},
  {Bradaschia}, {Brady}, {Bramley}, {Branchesi}, {Brau}, {Briant}, {Briggs},
  {Brighenti}, {Brillet}, {Brinkmann}, {Brisson}, {Brockill}, {Brooks},
  {Brown}, {Brunett}, {Buikema}, {Bulik}, {Bulten}, {Buonanno}, {Buskulic},
  {Bustamante Rosell}, {Buy}, {Byer}, {Cabero}, {Cadonati}, {Cagnoli},
  {Cahillane}, {Calder{\'o}n Bustillo}, {Callister}, {Calloni}, {Camp},
  {Campbell}, {Canepa}, {Cannon}, {Cao}, {Cao}, {Capocasa}, {Carbognani},
  {Caride}, {Carney}, {Carullo}, {Casanueva Diaz}, {Casentini}, {Caudill},
  {Cavagli{\`a}}, {Cavalier}, {Cavalieri}, {Cella}, {Cerd{\'a}-Dur{\'a}n},
  {Cerretani}, {Cesarini}, {Chaibi}, {Chakravarti}, {Chamberlin}, {Chan},
  {Chao}, {Charlton}, {Chase}, {Chassande-Mottin}, {Chatterjee}, {Chaturvedi},
  {Chatziioannou}, {Cheeseboro}, {Chen}, {Chen}, {Chen}, {Cheng}, {Cheong},
  {Chia}, {Chincarini}, {Chiummo}, {Cho}, {Cho}, {Cho}, {Christensen}, {Chu},
  {Chua}, {Chung}, {Chung}, {Ciani}, {Ciobanu}, {Ciolfi}, {Cipriano}, {Cirone},
  {Clara}, {Clark}, {Clearwater}, {Cleva}, {Cocchieri}, {Coccia}, {Cohadon},
  {Cohen}, {Colgan}, {Colleoni}, {Collette}, {Collins}, {Cominsky},
  {Constancio}, {Conti}, {Cooper}, {Corban}, {Corbitt}, {Cordero-Carri{\'o}n},
  {Corley}, {Cornish}, {Corsi}, {Cortese}, {Costa}, {Cotesta}, {Coughlin},
  {Coughlin}, {Coulon}, {Countryman}, {Couvares}, {Covas}, {Cowan}, {Coward},
  {Cowart}, {Coyne}, {Coyne}, {Creighton}, {Creighton}, {Cripe}, {Croquette},
  {Crowder}, {Cullen}, {Cumming}, {Cunningham}, {Cuoco}, {Canton}, {D{\'a}lya},
  {Danilishin}, {D'Antonio}, {Danzmann}, {Dasgupta}, {Da Silva Costa},
  {Datrier}, {Dattilo}, {Dave}, {Davier}, {Davis}, {Daw}, {DeBra},
  {Deenadayalan}, {Degallaix}, {De Laurentis}, {Del{\'e}glise}, {Del Pozzo},
  {DeMarchi}, {Demos}, {Dent}, {De Pietri}, {Derby}, {De Rosa}, {De Rossi},
  {DeSalvo}, {de Varona}, {Dhurandhar}, {D{\'\i}az}, {Dietrich}, {Di Fiore},
  {Di Giovanni}, {Di Girolamo}, {Di Lieto}, {Ding}, {Di Pace}, {Di Palma}, {Di
  Renzo}, {Dmitriev}, {Doctor}, {Donovan}, {Dooley}, {Doravari}, {Dorrington},
  {Downes}, {Drago}, {Driggers}, {Du}, {Ducoin}, {Dupej}, {Dwyer}, {Easter},
  {Edo}, {Edwards}, {Effler}, {Ehrens}, {Eichholz}, {Eikenberry}, {Eisenmann},
  {Eisenstein}, {Essick}, {Estelles}, {Estevez}, {Etienne}, {Etzel}, {Evans},
  {Evans}, {Fafone}, {Fair}, {Fairhurst}, {Fan}, {Farinon}, {Farr}, {Farr},
  {Fauchon-Jones}, {Favata}, {Fays}, {Fazio}, {Fee}, {Feicht}, {Fejer}, {Feng},
  {Fernandez-Galiana}, {Ferrante}, {Ferreira}, {Ferreira}, {Ferrini},
  {Fidecaro}, {Fiori}, {Fiorucci}, {Fishbach}, {Fisher}, {Fishner},
  {Fitz-Axen}, {Flaminio}, {Fletcher}, {Flynn}, {Fong}, {Font}, {Forsyth},
  {Fournier}, {Frasca}, {Frasconi}, {Frei}, {Freise}, {Frey}, {Frey},
  {Fritschel}, {Frolov}, {Fulda}, {Fyffe}, {Gabbard}, {Gadre}, {Gaebel},
  {Gair}, {Gammaitoni}, {Ganija}, {Gaonkar}, {Garcia},
  {Garc{\'\i}a-Quir{\'o}s}, {Garufi}, {Gateley}, {Gaudio}, {Gaur}, {Gayathri},
  {Gemme}, {Genin}, {Gennai}, {George}, {George}, {Gergely}, {Germain},
  {Ghonge}, {Ghosh}, {Ghosh}, {Ghosh}, {Giacomazzo}, {Giaime}, {Giardina},
  {Giazotto}, {Gill}, {Giordano}, {Glover}, {Godwin}, {Goetz}, {Goetz},
  {Goncharov}, {Gonz{\'a}lez}, {Gonzalez Castro}, {Gopakumar}, {Gorodetsky},
  {Gossan}, {Gosselin}, {Gouaty}, {Grado}, {Graef}, {Granata}, {Grant}, {Gras},
  {Grassia}, {Gray}, {Gray}, {Greco}, {Green}, {Green}, {Gretarsson}, {Groot},
  {Grote}, {Grunewald}, {Gruning}, {Guidi}, {Gulati}, {Guo}, {Gupta}, {Gupta},
  {Gustafson}, {Gustafson}, {Haegel}, {Halim}, {Hall}, {Hall}, {Hamilton},
  {Hammond}, {Haney}, {Hanke}, {Hanks}, {Hanna}, {Hannam}, {Hannuksela},
  {Hanson}, {Hardwick}, {Haris}, {Harms}, {Harry}, {Harry}, {Haster},
  {Haughian}, {Hayes}, {Healy}, {Heidmann}, {Heintze}, {Heitmann}, {Hello},
  {Hemming}, {Hendry}, {Heng}, {Hennig}, {Heptonstall}, {Hernandez Vivanco},
  {Heurs}, {Hild}, {Hinderer}, {Hoak}, {Hochheim}, {Hofman}, {Holgado},
  {Holland}, {Holt}, {Holz}, {Hopkins}, {Horst}, {Hough}, {Howell}, {Hoy},
  {Hreibi}, {Huang}, {Huerta}, {Huet}, {Hughey}, {Hulko}, {Husa}, {Huttner},
  {Huynh-Dinh}, {Idzkowski}, {Iess}, {Ingram}, {Inta}, {Intini}, {Irwin},
  {Isa}, {Isac}, {Isi}, {Iyer}, {Izumi}, {Jacqmin}, {Jadhav}, {Jani},
  {Janthalur}, {Jaranowski}, {Jenkins}, {Jiang}, {Johnson}, {Johnson-McDaniel},
  {Jones}, {Jones}, {Jones}, {Jonker}, {Ju}, {Junker}, {Kalaghatgi},
  {Kalogera}, {Kamai}, {Kandhasamy}, {Kang}, {Kanner}, {Kapadia}, {Karki},
  {Karvinen}, {Kashyap}, {Kasprzack}, {Katsanevas}, {Katsavounidis}, {Katzman},
  {Kaufer}, {Kawabe}, {Keerthana}, {K{\'e}f{\'e}lian}, {Keitel}, {Kennedy},
  {Key}, {Khalili}, {Khan}, {Khan}, {Khan}, {Khan}, {Khazanov}, {Khursheed},
  {Kijbunchoo}, {Kim}, {Kim}, {Kim}, {Kim}, {Kim}, {Kim}, {Kimball}, {King},
  {King}, {Kinley-Hanlon}, {Kirchhoff}, {Kissel}, {Kleybolte}, {Klika},
  {Klimenko}, {Knowles}, {Koch}, {Koehlenbeck}, {Koekoek}, {Koley},
  {Kondrashov}, {Kontos}, {Koper}, {Korobko}, {Korth}, {Kowalska}, {Kozak},
  {Kringel}, {Krishnendu}, {Kr{\'o}lak}, {Kuehn}, {Kumar}, {Kumar}, {Kumar},
  {Kumar}, {Kuo}, {Kutynia}, {Kwang}, {Lackey}, {Lai}, {Lam}, {Landry}, {Lane},
  {Lang}, {Lange}, {Lantz}, {Lanza}, {Lartaux-Vollard}, {Lasky}, {Laxen},
  {Lazzarini}, {Lazzaro}, {Leaci}, {Leavey}, {Lecoeuche}, {Lee}, {Lee}, {Lee},
  {Lee}, {Lee}, {Lee}, {Lehmann}, {Lenon}, {Leroy}, {Letendre}, {Levin}, {Li},
  {Li}, {Li}, {Li}, {Lin}, {Linde}, {Linker}, {Littenberg}, {Liu}, {Liu}, {Lo},
  {Lockerbie}, {London}, {Longo}, {Lorenzini}, {Loriette}, {Lormand},
  {Losurdo}, {Lough}, {Lousto}, {Lovelace}, {Lower}, {L{\"u}ck}, {Lumaca},
  {Lundgren}, {Lynch}, {Ma}, {Macas}, {Macfoy}, {MacInnis}, {Macleod},
  {Macquet}, {Maga{\~n}a-Sandoval}, {Maga{\~n}a Zertuche}, {Magee}, {Majorana},
  {Maksimovic}, {Malik}, {Man}, {Mandic}, {Mangano}, {Mansell}, {Manske},
  {Mantovani}, {Marchesoni}, {Marion}, {M{\'a}rka}, {M{\'a}rka}, {Markakis},
  {Markosyan}, {Markowitz}, {Maros}, {Marquina}, {Marsat}, {Martelli},
  {Martin}, {Martin}, {Martynov}, {Mason}, {Massera}, {Masserot}, {Massinger},
  {Masso-Reid}, {Mastrogiovanni}, {Matas}, {Matichard}, {Matone}, {Mavalvala},
  {Mazumder}, {McCann}, {McCarthy}, {McClelland}, {McCormick}, {McCuller},
  {McGuire}, {McIver}, {McManus}, {McRae}, {McWilliams}, {Meacher}, {Meadors},
  {Mehmet}, {Mehta}, {Meidam}, {Melatos}, {Mendell}, {Mercer}, {Mereni},
  {Merilh}, {Merzougui}, {Meshkov}, {Messenger}, {Messick}, {Metzdorff},
  {Meyers}, {Miao}, {Michel}, {Middleton}, {Mikhailov}, {Milano}, {Miller},
  {Miller}, {Millhouse}, {Mills}, {Milovich-Goff}, {Minazzoli}, {Minenkov},
  {Mishkin}, {Mishra}, {Mistry}, {Mitra}, {Mitrofanov}, {Mitselmakher},
  {Mittleman}, {Mo}, {Moffa}, {Mogushi}, {Mohapatra}, {Montani}, {Moore},
  {Moraru}, {Moreno}, {Morisaki}, {Mours}, {Mow-Lowry}, {Mukherjee},
  {Mukherjee}, {Mukherjee}, {Mukund}, {Mullavey}, {Munch}, {Mu{\~n}iz},
  {Muratore}, {Murray}, {Nagar}, {Nardecchia}, {Naticchioni}, {Nayak},
  {Neilson}, {Nelemans}, {Nelson}, {Nery}, {Neunzert}, {Ng}, {Ng}, {Nguyen},
  {Nichols}, {Nielsen}, {Nissanke}, {Nitz}, {Nocera}, {North}, {Nuttall},
  {Obergaulinger}, {Oberling}, {O'Brien}, {O'Dea}, {Ogin}, {Oh}, {Oh}, {Ohme},
  {Ohta}, {Okada}, {Oliver}, {Oppermann}, {Oram}, {O'Reilly}, {Ormiston},
  {Ortega}, {O'Shaughnessy}, {Ossokine}, {Ottaway}, {Overmier}, {Owen}, {Pace},
  {Pagano}, {Page}, {Pai}, {Pai}, {Palamos}, {Palashov}, {Palomba},
  {Pal-Singh}, {Pan}, {Pang}, {Pang}, {Pankow}, {Pannarale}, {Pant},
  {Paoletti}, {Paoli}, {Papa}, {Parida}, {Parker}, {Pascucci}, {Pasqualetti},
  {Passaquieti}, {Passuello}, {Patil}, {Patricelli}, {Pearlstone}, {Pedersen},
  {Pedraza}, {Pedurand}, {Pele}, {Penn}, {Perego}, {Perez}, {Perreca},
  {Pfeiffer}, {Phelps}, {Phukon}, {Piccinni}, {Pichot}, {Piergiovanni},
  {Pillant}, {Pinard}, {Pirello}, {Pitkin}, {Poggiani}, {Pong}, {Ponrathnam},
  {Popolizio}, {Porter}, {Powell}, {Prajapati}, {Prasad}, {Prasai}, {Prasanna},
  {Pratten}, {Prestegard}, {Privitera}, {Prodi}, {Prokhorov}, {Puncken},
  {Punturo}, {Puppo}, {P{\"u}rrer}, {Qi}, {Quetschke}, {Quinonez}, {Quintero},
  {Quitzow-James}, {Raab}, {Radkins}, {Radulescu}, {Raffai}, {Raja}, {Rajan},
  {Rajbhandari}, {Rakhmanov}, {Ramirez}, {Ramos-Buades}, {Rana}, {Rao},
  {Rapagnani}, {Raymond}, {Razzano}, {Read}, {Regimbau}, {Rei}, {Reid},
  {Reitze}, {Ren}, {Ricci}, {Richardson}, {Richardson}, {Ricker},
  {Riemenschneider}, {Riles}, {Rizzo}, {Robertson}, {Robie}, {Robinet},
  {Rocchi}, {Rolland}, {Rollins}, {Roma}, {Romanelli}, {Romano}, {Romel},
  {Romie}, {Rose}, {Rosi{\'n}ska}, {Rosofsky}, {Ross}, {Rowan}, {R{\"u}diger},
  {Ruggi}, {Rutins}, {Ryan}, {Sachdev}, {Sadecki}, {Sakellariadou}, {Salafia},
  {Salconi}, {Saleem}, {Salemi}, {Samajdar}, {Sammut}, {Sanchez}, {Sanchez},
  {Sanchis-Gual}, {Sandberg}, {Sanders}, {Santiago}, {Sarin}, {Sassolas},
  {Sathyaprakash}, {Saulson}, {Sauter}, {Savage}, {Schale}, {Scheel},
  {Scheuer}, {Schmidt}, {Schnabel}, {Schofield}, {Sch{\"o}nbeck}, {Schreiber},
  {Schulte}, {Schutz}, {Schwalbe}, {Scott}, {Scott}, {Seidel}, {Sellers},
  {Sengupta}, {Sennett}, {Sentenac}, {Sequino}, {Sergeev}, {Setyawati},
  {Shaddock}, {Shaffer}, {Shahriar}, {Shaner}, {Shao}, {Sharma}, {Shawhan},
  {Shen}, {Shink}, {Shoemaker}, {Shoemaker}, {ShyamSundar}, {Siellez},
  {Sieniawska}, {Sigg}, {Silva}, {Singer}, {Singh}, {Singhal}, {Sintes},
  {Sitmukhambetov}, {Skliris}, {Slagmolen}, {Slaven-Blair}, {Smith}, {Smith},
  {Somala}, {Son}, {Sorazu}, {Sorrentino}, {Souradeep}, {Sowell}, {Spencer},
  {Srivastava}, {Srivastava}, {Staats}, {Stachie}, {Standke}, {Steer},
  {Steinke}, {Steinlechner}, {Steinlechner}, {Steinmeyer}, {Stevenson},
  {Stocks}, {Stone}, {Stops}, {Strain}, {Stratta}, {Strigin}, {Strunk},
  {Sturani}, {Stuver}, {Sudhir}, {Summerscales}, {Sun}, {Sunil}, {Suresh},
  {Sutton}, {Swinkels}, {Szczepa{\'n}czyk}, {Tacca}, {Tait}, {Talbot},
  {Talukder}, {Tanner}, {T{\'a}pai}, {Taracchini}, {Tasson}, {Taylor}, {Thies},
  {Thomas}, {Thomas}, {Thondapu}, {Thorne}, {Thrane}, {Tiwari}, {Tiwari},
  {Tiwari}, {Toland}, {Tonelli}, {Tornasi}, {Torres-Forn{\'e}}, {Torrie},
  {T{\"o}yr{\"a}}, {Travasso}, {Traylor}, {Tringali}, {Trovato}, {Trozzo},
  {Trudeau}, {Tsang}, {Tse}, {Tso}, {Tsukada}, {Tsuna}, {Tuyenbayev}, {Ueno},
  {Ugolini}, {Unnikrishnan}, {Urban}, {Usman}, {Vahlbruch}, {Vajente},
  {Valdes}, {van Bakel}, {van Beuzekom}, {van den Brand}, {Van Den Broeck},
  {Vander-Hyde}, {van Heijningen}, {van der Schaaf}, {van Veggel}, {Vardaro},
  {Varma}, {Vass}, {Vas{\'u}th}, {Vecchio}, {Vedovato}, {Veitch}, {Veitch},
  {Venkateswara}, {Venugopalan}, {Verkindt}, {Vetrano}, {Vicer{\'e}}, {Viets},
  {Vine}, {Vinet}, {Vitale}, {Vo}, {Vocca}, {Vorvick}, {Vyatchanin}, {Wade},
  {Wade}, {Wade}, {Walet}, {Walker}, {Wallace}, {Walsh}, {Wang}, {Wang},
  {Wang}, {Wang}, {Wang}, {Ward}, {Warden}, {Warner}, {Was}, {Watchi},
  {Weaver}, {Wei}, {Weinert}, {Weinstein}, {Weiss}, {Wellmann}, {Wen},
  {Wessel}, {We{\ss}els}, {Westhouse}, {Wette}, {Whelan}, {White}, {Whiting},
  {Whittle}, {Wilken}, {Williams}, {Williamson}, {Willis}, {Willke}, {Wimmer},
  {Winkler}, {Wipf}, {Wittel}, {Woan}, {Woehler}, {Wofford}, {Worden},
  {Wright}, {Wu}, {Wysocki}, {Xiao}, {Yamamoto}, {Yancey}, {Yang}, {Yap},
  {Yazback}, {Yeeles}, {Yu}, {Yu}, {Yuen}, {Yvert}, {Zadro{\.Z}ny}, {Zanolin},
  {Zappa}, {Zelenova}, {Zendri}, {Zevin}, {Zhang}, {Zhang}, {Zhang}, {Zhao},
  {Zhou}, {Zhou}, {Zhu}, {Zimmerman}, {Zlochower}, {Zucker}, {Zweizig}, {LIGO
  Scientific Collaboration}, \& {Virgo Collaboration}}]{2019PhRvX...9c1040A}
{Abbott}, B.~P., {Abbott}, R., {Abbott}, T.~D., {$et~al$.} 2019, Physical
  Review X, 9, 031040

\bibitem[{{Abbott} {$et~al$.}(2020){Abbott}, {Abbott}, {Abraham}, {Acernese},
  {Ackley}, {Adams}, {Adams}, {Adhikari}, {Adya}, {Affeldt}, {Agathos},
  {Agatsuma}, {Aggarwal}, {Aguiar}, {Aiello}, {Ain}, {Ajith}, {Akcay}, {Allen},
  {Allocca}, {Altin}, {Amato}, {Anand}, {Ananyeva}, {Anderson}, {Anderson},
  {Angelova}, {Ansoldi}, {Antelis}, {Antier}, {Appert}, {Arai}, {Araya},
  {Areeda}, {Ar{\`e}ne}, {Arnaud}, {Aronson}, {Arun}, {Asali}, {Ascenzi},
  {Ashton}, {Aston}, {Astone}, {Aubin}, {Aufmuth}, {AultONeal}, {Austin},
  {Avendano}, {Babak}, {Badaracco}, {Bader}, {Bae}, {Baer}, {Bagnasco},
  {Baird}, {Ball}, {Ballardin}, {Ballmer}, {Bals}, {Balsamo}, {Baltus},
  {Banagiri}, {Bankar}, {Bankar}, {Barayoga}, {Barbieri}, {Barish}, {Barker},
  {Barneo}, {Barnum}, {Barone}, {Barr}, {Barsotti}, {Barsuglia}, {Barta},
  {Bartlett}, {Bartos}, {Bassiri}, {Basti}, {Bawaj}, {Bayley}, {Bazzan},
  {Becher}, {B{\'e}csy}, {Bedakihale}, {Bejger}, {Belahcene}, {Beniwal},
  {Benjamin}, {Bennett}, {Bentley}, {Bergamin}, {Berger}, {Bergmann},
  {Bernuzzi}, {Berry}, {Bersanetti}, {Bertolini}, {Betzwieser}, {Bhandare},
  {Bhandari}, {Bhattacharjee}, {Bidler}, {Bilenko}, {Billingsley}, {Birney},
  {Birnholtz}, {Biscans}, {Bischi}, {Biscoveanu}, {Bisht}, {Bitossi},
  {Bizouard}, {Blackburn}, {Blackman}, {Blair}, {Blair}, {Blair}, {Blanch},
  {Bobba}, {Bode}, {Boer}, {Boetzel}, {Bogaert}, {Boldrini}, {Bondu},
  {Bonnand}, {Bonilla}, {Booker}, {Boom}, {Bork}, {Boschi}, {Bose},
  {Bossilkov}, {Boudart}, {Bouffanais}, {Bozzi}, {Bradaschia}, {Brady},
  {Bramley}, {Branchesi}, {Brau}, {Breschi}, {Briant}, {Briggs}, {Brighenti},
  {Brillet}, {Brinkmann}, {Brockill}, {Brooks}, {Brooks}, {Brown}, {Brunett},
  {Bruno}, {Bruntz}, {Buikema}, {Bulik}, {Bulten}, {Buonanno}, {Buscicchio},
  {Buskulic}, {Byer}, {Cabero}, {Cadonati}, {Caesar}, {Cagnoli}, {Cahillane},
  {Calder{\'o}n Bustillo}, {Callaghan}, {Callister}, {Calloni}, {Camp},
  {Canepa}, {Cannon}, {Cao}, {Cao}, {Carapella}, {Carbognani}, {Carney},
  {Carpinelli}, {Carullo}, {Carver}, {Casanueva Diaz}, {Casentini}, {Caudill},
  {Cavagli{\`a}}, {Cavalier}, {Cavalieri}, {Cella}, {Cerd{\'a}-Dur{\'a}n},
  {Cesarini}, {Chaibi}, {Chakravarti}, {Chan}, {Chan}, {Chandra}, {Chanial},
  {Chao}, {Charlton}, {Chase}, {Chassande-Mottin}, {Chatterjee},
  {Chattopadhyay}, {Chaturvedi}, {Chatziioannou}, {Chen}, {Chen}, {Chen},
  {Chen}, {Cheng}, {Cheong}, {Chia}, {Chiadini}, {Chierici}, {Chincarini},
  {Chiummo}, {Cho}, {Cho}, {Cho}, {Choate}, {Christensen}, {Chu}, {Chua},
  {Chung}, {Chung}, {Ciani}, {Ciecielag}, {Cie{\'s}lar}, {Cifaldi}, {Ciobanu},
  {Ciolfi}, {Cipriano}, {Cirone}, {Clara}, {Clark}, {Clark}, {Clarke},
  {Clearwater}, {Clesse}, {Cleva}, {Coccia}, {Cohadon}, {Cohen}, {Colleoni},
  {Collette}, {Collins}, {Colpi}, {Constancio}, {Conti}, {Cooper}, {Corban},
  {Corbitt}, {Cordero-Carri{\'o}n}, {Corezzi}, {Corley}, {Cornish}, {Corre},
  {Corsi}, {Cortese}, {Costa}, {Cotesta}, {Coughlin}, {Coughlin}, {Coulon},
  {Countryman}, {Cousins}, {Couvares}, {Covas}, {Coward}, {Cowart}, {Coyne},
  {Coyne}, {Creighton}, {Creighton}, {Croquette}, {Crowder}, {Cudell},
  {Cullen}, {Cumming}, {Cummings}, {Cunningham}, {Cuoco}, {Curylo}, {Dal
  Canton}, {D{\'a}lya}, {Dana}, {DaneshgaranBajastani}, {D'Angelo}, {Danila},
  {Danilishin}, {D'Antonio}, {Danzmann}, {Darsow-Fromm}, {Dasgupta}, {Datrier},
  {Dattilo}, {Dave}, {Davier}, {Davies}, {Davis}, {Daw}, {Dean}, {DeBra},
  {Deenadayalan}, {Degallaix}, {De Laurentis}, {Del{\'e}glise}, {Del Favero},
  {De Lillo}, {De Lillo}, {Del Pozzo}, {DeMarchi}, {De Matteis}, {D'Emilio},
  {Demos}, {Denker}, {Dent}, {Depasse}, {De Pietri}, {De Rosa}, {De Rossi},
  {DeSalvo}, {de Varona}, {Dhurandhar}, {D{\'\i}az}, {Diaz-Ortiz}, {Didio},
  {Dietrich}, {Di Fiore}, {DiFronzo}, {Di Giorgio}, {Di Giovanni}, {Di
  Giovanni}, {Di Girolamo}, {Di Lieto}, {Ding}, {Di Pace}, {Di Palma}, {Di
  Renzo}, {Divakarla}, {Dmitriev}, {Doctor}, {D'Onofrio}, {Donovan}, {Dooley},
  {Doravari}, {Dorrington}, {Downes}, {Drago}, {Driggers}, {Du}, {Ducoin},
  {Dupej}, {Durante}, {D'Urso}, {Duverne}, {Dwyer}, {Easter}, {Eddolls},
  {Edelman}, {Edo}, {Edy}, {Effler}, {Eichholz}, {Eikenberry}, {Eisenmann},
  {Eisenstein}, {Ejlli}, {Errico}, {Essick}, {Estell{\'e}s}, {Estevez},
  {Etienne}, {Etzel}, {Evans}, {Evans}, {Ewing}, {Fafone}, {Fair}, {Fairhurst},
  {Fan}, {Farah}, {Farinon}, {Farr}, {Farr}, {Fauchon-Jones}, {Favata}, {Fays},
  {Fazio}, {Feicht}, {Fejer}, {Feng}, {Fenyvesi}, {Ferguson},
  {Fernandez-Galiana}, {Ferrante}, {Ferreira}, {Fidecaro}, {Figura}, {Fiori},
  {Fiorucci}, {Fishbach}, {Fisher}, {Fishner}, {Fittipaldi}, {Fitz-Axen},
  {Fiumara}, {Flaminio}, {Floden}, {Flynn}, {Fong}, {Font}, {Forsyth},
  {Fournier}, {Frasca}, {Frasconi}, {Frei}, {Freise}, {Frey}, {Frey},
  {Fritschel}, {Frolov}, {Fronz{\'e}}, {Fulda}, {Fyffe}, {Gabbard}, {Gadre},
  {Gaebel}, {Gair}, {Gais}, {Galaudage}, {Gamba}, {Ganapathy}, {Ganguly},
  {Gaonkar}, {Garaventa}, {Garc{\'\i}a-Quir{\'o}s}, {Garufi}, {Gateley},
  {Gaudio}, {Gayathri}, {Gemme}, {Gennai}, {George}, {George}, {George},
  {Gergely}, {Ghonge}, {Ghosh}, {Ghosh}, {Ghosh}, {Giacomazzo}, {Giacoppo},
  {Giaime}, {Giardina}, {Gibson}, {Gier}, {Gill}, {Giri}, {Glanzer}, {Gleckl},
  {Godwin}, {Goetz}, {Goetz}, {Gohlke}, {Goncharov}, {Gonz{\'a}lez},
  {Gopakumar}, {Gossan}, {Gosselin}, {Gouaty}, {Grace}, {Grado}, {Granata},
  {Granata}, {Grant}, {Gras}, {Grassia}, {Gray}, {Gray}, {Greco}, {Green},
  {Green}, {Gretarsson}, {Griggs}, {Grignani}, {Grimaldi}, {Grimes}, {Grimm},
  {Grote}, {Grunewald}, {Gruning}, {Guerrero}, {Guidi}, {Guimaraes},
  {Guix{\'e}}, {Gulati}, {Guo}, {Gupta}, {Gupta}, {Gupta}, {Gustafson},
  {Gustafson}, {Guzman}, {Haegel}, {Halim}, {Hall}, {Hamilton}, {Hammond},
  {Haney}, {Hanke}, {Hanks}, {Hanna}, {Hannam}, {Hannuksela}, {Hannuksela},
  {Hansen}, {Hansen}, {Hanson}, {Harder}, {Hardwick}, {Haris}, {Harms},
  {Harry}, {Harry}, {Hartwig}, {Hasskew}, {Haster}, {Haughian}, {Hayes},
  {Healy}, {Heidmann}, {Heintze}, {Heinze}, {Heinzel}, {Heitmann}, {Hellman},
  {Hello}, {Helmling-Cornell}, {Hemming}, {Hendry}, {Heng}, {Hennes}, {Hennig},
  {Hennig}, {Hernandez Vivanco}, {Heurs}, {Hild}, {Hill}, {Hines}, {Hochheim},
  {Hofgard}, {Hofman}, {Hohmann}, {Holgado}, {Holland}, {Hollows}, {Holmes},
  {Holt}, {Holz}, {Hopkins}, {Horst}, {Hough}, {Howell}, {Hoy}, {Hoyland},
  {Huang}, {H{\"u}bner}, {Huddart}, {Huerta}, {Hughey}, {Hui}, {Husa},
  {Huttner}, {Hutzler}, {Huxford}, {Huynh-Dinh}, {Idzkowski}, {Iess},
  {Imperato}, {Inchauspe}, {Ingram}, {Intini}, {Isi}, {Iyer},
  {JaberianHamedan}, {Jacqmin}, {Jadhav}, {Jadhav}, {James}, {Jani},
  {Janssens}, {Janthalur}, {Jaranowski}, {Jariwala}, {Jaume}, {Jenkins},
  {Jeunon}, {Jiang}, {Johns}, {Johnson-McDaniel}, {Jones}, {Jones}, {Jones},
  {Jones}, {Jones}, {Jonker}, {Ju}, {Junker}, {Kalaghatgi}, {Kalogera},
  {Kamai}, {Kandhasamy}, {Kang}, {Kanner}, {Kapadia}, {Kapasi}, {Karathanasis},
  {Karki}, {Kashyap}, {Kasprzack}, {Kastaun}, {Katsanevas}, {Katsavounidis},
  {Katzman}, {Kawabe}, {K{\'e}f{\'e}lian}, {Keitel}, {Key}, {Khadka},
  {Khalili}, {Khan}, {Khan}, {Khazanov}, {Khetan}, {Khursheed}, {Kijbunchoo},
  {Kim}, {Kim}, {Kim}, {Kim}, {Kim}, {Kim}, {Kimball}, {King}, {Kinley-Hanlon},
  {Kirchhoff}, {Kissel}, {Kleybolte}, {Klimenko}, {Knowles}, {Knyazev}, {Koch},
  {Koehlenbeck}, {Koekoek}, {Koley}, {Kolstein}, {Komori}, {Kondrashov},
  {Kontos}, {Koper}, {Korobko}, {Korth}, {Kovalam}, {Kozak}, {Kr{\"a}mer},
  {Kringel}, {Krishnendu}, {Kr{\'o}lak}, {Kuehn}, {Kumar}, {Kumar}, {Kumar},
  {Kumar}, {Kuns}, {Kwang}, {Lackey}, {Laghi}, {Lalande}, {Lam}, {Lamberts},
  {Landry}, {Lane}, {Lang}, {Lange}, {Lantz}, {Lanza}, {La Rosa},
  {Lartaux-Vollard}, {Lasky}, {Laxen}, {Lazzarini}, {Lazzaro}, {Leaci},
  {Leavey}, {Lecoeuche}, {Lee}, {Lee}, {Lee}, {Lee}, {Lehmann}, {Leon},
  {Leroy}, {Letendre}, {Levin}, {Li}, {Li}, {Li}, {Li}, {Li}, {Linde},
  {Linker}, {Linley}, {Littenberg}, {Liu}, {Liu}, {Llorens-Monteagudo}, {Lo},
  {Lockwood}, {London}, {Longo}, {Lorenzini}, {Loriette}, {Lormand}, {Losurdo},
  {Lough}, {Lousto}, {Lovelace}, {L{\"u}ck}, {Lumaca}, {Lundgren}, {Ma},
  {Macas}, {MacInnis}, {Macleod}, {MacMillan}, {Macquet}, {Maga{\~n}a
  Hernandez}, {Maga{\~n}a-Sandoval}, {Magazz{\`u}}, {Magee}, {Majorana},
  {Maksimovic}, {Maliakal}, {Malik}, {Man}, {Mandic}, {Mangano}, {Mansell},
  {Manske}, {Mantovani}, {Mapelli}, {Marchesoni}, {Marion}, {M{\'a}rka},
  {M{\'a}rka}, {Markakis}, {Markosyan}, {Markowitz}, {Maros}, {Marquina},
  {Marsat}, {Martelli}, {Martin}, {Martin}, {Martinez}, {Martinez}, {Martynov},
  {Masalehdan}, {Mason}, {Massera}, {Masserot}, {Massinger}, {Masso-Reid},
  {Mastrogiovanni}, {Matas}, {Mateu-Lucena}, {Matichard}, {Matiushechkina},
  {Mavalvala}, {Maynard}, {McCann}, {McCarthy}, {McClelland}, {McCormick},
  {McCuller}, {McGuire}, {McIsaac}, {McIver}, {McManus}, {McRae}, {McWilliams},
  {Meacher}, {Meadors}, {Mehmet}, {Mehta}, {Melatos}, {Melchor}, {Mendell},
  {Menendez-Vazquez}, {Mercer}, {Mereni}, {Merfeld}, {Merilh}, {Merritt},
  {Merzougui}, {Meshkov}, {Messenger}, {Messick}, {Metzdorff}, {Meyers},
  {Meylahn}, {Mhaske}, {Miani}, {Miao}, {Michaloliakos}, {Michel}, {Middleton},
  {Milano}, {Miller}, {Millhouse}, {Mills}, {Milotti}, {Milovich-Goff},
  {Minazzoli}, {Minenkov}, {Mir}, {Mishkin}, {Mishra}, {Mistry}, {Mitra},
  {Mitrofanov}, {Mitselmakher}, {Mittleman}, {Mo}, {Mogushi}, {Mohapatra},
  {Mohite}, {Molina}, {Molina-Ruiz}, {Mondin}, {Montani}, {Moore}, {Moraru},
  {Morawski}, {Moreno}, {Morisaki}, {Mours}, {Mow-Lowry}, {Mozzon},
  {Muciaccia}, {Mukherjee}, {Mukherjee}, {Mukherjee}, {Mukherjee}, {Mukund},
  {Mullavey}, {Munch}, {Mu{\~n}iz}, {Murray}, {Nadji}, {Nagar}, {Nardecchia},
  {Naticchioni}, {Nayak}, {Neil}, {Neilson}, {Nelemans}, {Nelson}, {Nery},
  {Neunzert}, {Nitz}, {Ng}, {Ng}, {Nguyen}, {Nguyen}, {Nguyen}, {Nichols},
  {Nissanke}, {Nocera}, {Noh}, {North}, {Nothard}, {Nuttall}, {Oberling},
  {O'Brien}, {O'Dell}, {Oganesyan}, {Ogin}, {Oh}, {Oh}, {Ohme}, {Ohta},
  {Okada}, {Olivetto}, {Oppermann}, {Oram}, {O'Reilly}, {Ormiston}, {Ortega},
  {O'Shaughnessy}, {Ossokine}, {Osthelder}, {Ottaway}, {Overmier}, {Owen},
  {Pace}, {Pagano}, {Page}, {Pagliaroli}, {Pai}, {Pai}, {Palamos}, {Palashov},
  {Palomba}, {Pan}, {Panda}, {Pang}, {Pankow}, {Pannarale}, {Pant}, {Paoletti},
  {Paoli}, {Paolone}, {Parker}, {Pascucci}, {Pasqualetti}, {Passaquieti},
  {Passuello}, {Patel}, {Patricelli}, {Payne}, {Pechsiri}, {Pedraza},
  {Pegoraro}, {Pele}, {Penn}, {Perego}, {Perez}, {P{\'e}rigois}, {Perreca},
  {Perri{\`e}s}, {Petermann}, {Petterson}, {Pfeiffer}, {Pham}, {Phukon},
  {Piccinni}, {Pichot}, {Piendibene}, {Piergiovanni}, {Pierini}, {Pierro},
  {Pillant}, {Pilo}, {Pinard}, {Pinto}, {Piotrzkowski}, {Pirello}, {Pitkin},
  {Placidi}, {Plastino}, {Pluchar}, {Poggiani}, {Polini}, {Pong}, {Ponrathnam},
  {Popolizio}, {Porter}, {Poverman}, {Powell}, {Pracchia}, {Prajapati},
  {Prasai}, {Prasanna}, {Pratten}, {Prestegard}, {Principe}, {Prodi},
  {Prokhorov}, {Prosposito}, {Prudenzi}, {Puecher}, {Punturo}, {Puosi},
  {Puppo}, {P{\"u}rrer}, {Qi}, {Quetschke}, {Quinonez}, {Quitzow-James},
  {Raab}, {Raaijmakers}, {Radkins}, {Radulesco}, {Raffai}, {Rafferty}, {Rail},
  {Raja}, {Rajan}, {Rajbhandari}, {Rakhmanov}, {Ramirez}, {Ramirez},
  {Ramos-Buades}, {Rana}, {Rao}, {Rapagnani}, {Rapol}, {Ratto}, {Raymond},
  {Razzano}, {Read}, {Regimbau}, {Rei}, {Reid}, {Reitze}, {Rettegno}, {Ricci},
  {Richardson}, {Richardson}, {Richardson}, {Ricker}, {Riemenschneider},
  {Riles}, {Rizzo}, {Robertson}, {Robinet}, {Rocchi}, {Rocha}, {Rodriguez},
  {Rodriguez-Soto}, {Rolland}, {Rollins}, {Roma}, {Romanelli}, {Romano},
  {Romel}, {Romero}, {Romero-Shaw}, {Romie}, {Ronchini}, {Rose}, {Rose},
  {Rose}, {Rosell}, {Rosi{\'n}ska}, {Rosofsky}, {Ross}, {Rowan}, {Rowlinson},
  {Roy}, {Roy}, {Ruggi}, {Ryan}, {Sachdev}, {Sadecki}, {Sadiq},
  {Sakellariadou}, {Salafia}, {Salconi}, {Saleem}, {Samajdar}, {Sanchez},
  {Sanchez}, {Sanchez}, {Sanchis-Gual}, {Sanders}, {Sandles}, {Santiago},
  {Santos}, {Saravanan}, {Sarin}, {Sassolas}, {Sathyaprakash}, {Sauter},
  {Savage}, {Savant}, {Sawant}, {Sayah}, {Schaetzl}, {Schale}, {Scheel},
  {Scheuer}, {Schindler-Tyka}, {Schmidt}, {Schnabel}, {Schofield},
  {Sch{\"o}nbeck}, {Schreiber}, {Schulte}, {Schutz}, {Schwarm}, {Schwartz},
  {Scott}, {Scott}, {Seglar-Arroyo}, {Seidel}, {Sellers}, {Sengupta},
  {Sennett}, {Sentenac}, {Sequino}, {Sergeev}, {Setyawati}, {Shaffer},
  {Shahriar}, {Sharifi}, {Sharma}, {Sharma}, {Shawhan}, {Shen}, {Shikauchi},
  {Shink}, {Shoemaker}, {Shoemaker}, {Shukla}, {ShyamSundar}, {Sieniawska},
  {Sigg}, {Singer}, {Singh}, {Singh}, {Singha}, {Singhal}, {Sintes}, {Sipala},
  {Skliris}, {Slagmolen}, {Slaven-Blair}, {Smetana}, {Smith}, {Smith},
  {Somala}, {Son}, {Soni}, {Soni}, {Sorazu}, {Sordini}, {Sorrentino},
  {Sorrentino}, {Soulard}, {Souradeep}, {Sowell}, {Spencer}, {Spera},
  {Srivastava}, {Srivastava}, {Staats}, {Stachie}, {Steer}, {Steinhoff},
  {Steinke}, {Steinlechner}, {Steinlechner}, {Steinmeyer}, {Stevenson},
  {Stolle-McAllister}, {Stops}, {Stover}, {Strain}, {Stratta}, {Strunk},
  {Sturani}, {Stuver}, {S{\"u}dbeck}, {Sudhagar}, {Sudhir}, {Suh},
  {Summerscales}, {Sun}, {Sun}, {Sunil}, {Sur}, {Suresh}, {Sutton}, {Swinkels},
  {Szczepa{\'n}czyk}, {Tacca}, {Tait}, {Talbot}, {Tanasijczuk}, {Tanner},
  {Tao}, {Tapia}, {Tapia San Martin}, {Tasson}, {Taylor}, {Tenorio},
  {Terkowski}, {Thirugnanasambandam}, {Thomas}, {Thomas}, {Thomas}, {Thompson},
  {Thondapu}, {Thorne}, {Thrane}, {Tiwari}, {Tiwari}, {Tiwari}, {Toland},
  {Tolley}, {Tonelli}, {Tornasi}, {Torres-Forn{\'e}}, {Torrie}, {Melo},
  {T{\"o}yr{\"a}}, {Tran}, {Trapananti}, {Travasso}, {Traylor}, {Tringali},
  {Tripathee}, {Trovato}, {Trudeau}, {Tsai}, {Tsang}, {Tse}, {Tso}, {Tsukada},
  {Tsuna}, {Tsutsui}, {Turconi}, {Ubhi}, {Udall}, {Ueno}, {Ugolini},
  {Unnikrishnan}, {Urban}, {Usman}, {Utina}, {Vahlbruch}, {Vajente}, {Vajpeyi},
  {Valdes}, {Valentini}, {Valsan}, {van Bakel}, {van Beuzekom}, {van den
  Brand}, {Van Den Broeck}, {Vander-Hyde}, {van der Schaaf}, {van Heijningen},
  {Vardaro}, {Vargas}, {Varma}, {Vass}, {Vas{\'u}th}, {Vecchio}, {Vedovato},
  {Veitch}, {Veitch}, {Venkateswara}, {Venneberg}, {Venugopalan}, {Verkindt},
  {Verma}, {Veske}, {Vetrano}, {Vicer{\'e}}, {Viets}, {Vijaykumar},
  {Villa-Ortega}, {Vinet}, {Vitale}, {Vo}, {Vocca}, {Vorvick}, {Vyatchanin},
  {Wade}, {Wade}, {Wade}, {Walet}, {Walker}, {Wallace}, {Wallace}, {Walsh},
  {Wang}, {Wang}, {Wang}, {Wang}, {Ward}, {Warner}, {Was}, {Washington},
  {Watchi}, {Weaver}, {Wei}, {Weinert}, {Weinstein}, {Weiss}, {Wellmann},
  {Wen}, {We{\ss}els}, {Westhouse}, {Wette}, {Whelan}, {White}, {White},
  {Whiting}, {Whittle}, {Wilken}, {Williams}, {Williams}, {Williamson},
  {Willis}, {Willke}, {Wilson}, {Wimmer}, {Winkler}, {Wipf}, {Woan}, {Woehler},
  {Wofford}, {Wong}, {Wrangel}, {Wright}, {Wu}, {Wysocki}, {Xiao}, {Yamamoto},
  {Yang}, {Yang}, {Yang}, {Yap}, {Yeeles}, {Yoon}, {Yu}, {Yu}, {Yuen},
  {Zadro{\.z}ny}, {Zanolin}, {Zelenova}, {Zendri}, {Zevin}, {Zhang}, {Zhang},
  {Zhang}, {Zhang}, {Zhao}, {Zhao}, {Zheng}, {Zhou}, {Zhou}, {Zhu},
  {Zimmerman}, {Zlochower}, {Zucker}, \& {Zweizig}}]{2020arXiv201014527A}
{Abbott}, R., {Abbott}, T.~D., {Abraham}, S., {$et~al$.} 2020, arXiv e-prints,
  arXiv:2010.14527

\bibitem[{{Abdikamalov} {$et~al$.}(2014){Abdikamalov}, {Gossan}, {DeMaio}, \&
  {Ott}}]{2014PhRvD..90d4001A}
{Abdikamalov}, E., {Gossan}, S., {DeMaio}, A.~M., \& {Ott}, C.~D. 2014, \prd,
  90, 044001

\bibitem[{{Adams} {$et~al$.}(2013){Adams}, {Kochanek}, {Beacom}, {Vagins}, \&
  {Stanek}}]{2013ApJ...778..164A}
{Adams}, S.~M., {Kochanek}, C.~S., {Beacom}, J.~F., {Vagins}, M.~R., \&
  {Stanek}, K.~Z. 2013, \apj, 778, 164

\bibitem[{{Afle} \& {Brown}(2021)}]{2021PhRvD.103b3005A}
{Afle}, C., \& {Brown}, D.~A. 2021, \prd, 103, 023005

\bibitem[{{Andresen} {$et~al$.}(2021){Andresen}, {Glas}, \&
  {Janka}}]{2021MNRAS.503.3552A}
{Andresen}, H., {Glas}, R., \& {Janka}, H.~T. 2021, \mnras, 503, 3552

\bibitem[{{Andresen} {$et~al$.}(2017){Andresen}, {M{\"u}ller}, {M{\"u}ller}, \&
  {Janka}}]{2017MNRAS.468.2032A}
{Andresen}, H., {M{\"u}ller}, B., {M{\"u}ller}, E., \& {Janka}, H.~T. 2017,
  \mnras, 468, 2032

\bibitem[{{Andresen} {$et~al$.}(2019){Andresen}, {M{\"u}ller}, {Janka},
  {Summa}, {Gill}, \& {Zanolin}}]{2019MNRAS.486.2238A}
{Andresen}, H., {M{\"u}ller}, E., {Janka}, H.~T., {$et~al$.} 2019, \mnras, 486,
  2238

\bibitem[{{Bennett}(2005)}]{2005ApJ...633..906B}
{Bennett}, D.~P. 2005, \apj, 633, 906

\bibitem[{{Bozzetto} {$et~al$.}(2017){Bozzetto}, {Filipovi{\'c}},
  {Vukoti{\'c}}, {Pavlovi{\'c}}, {Uro{\v{s}}evi{\'c}}, {Kavanagh}, {Arbutina},
  {Maggi}, {Sasaki}, {Haberl}, {Crawford}, {Roper}, {Grieve}, \&
  {Points}}]{2017ApJS..230....2B}
{Bozzetto}, L.~M., {Filipovi{\'c}}, M.~D., {Vukoti{\'c}}, B., {$et~al$.} 2017,
  \apjs, 230, 2

\bibitem[{{Broadhurst} {$et~al$.}(2018){Broadhurst}, {Diego}, \&
  {Smoot}}]{2018arXiv180205273B}
{Broadhurst}, T., {Diego}, J.~M., \& {Smoot}, George, I. 2018, arXiv e-prints,
  arXiv:1802.05273

\bibitem[{{Cerd{\'a}-Dur{\'a}n} {$et~al$.}(2013){Cerd{\'a}-Dur{\'a}n},
  {DeBrye}, {Aloy}, {Font}, \& {Obergaulinger}}]{2013ApJ...779L..18C}
{Cerd{\'a}-Dur{\'a}n}, P., {DeBrye}, N., {Aloy}, M.~A., {Font}, J.~A., \&
  {Obergaulinger}, M. 2013, \apjl, 779, L18

\bibitem[{{Christian} {$et~al$.}(2018){Christian}, {Vitale}, \&
  {Loeb}}]{2018PhRvD..98j3022C}
{Christian}, P., {Vitale}, S., \& {Loeb}, A. 2018, \prd, 98, 103022

\bibitem[{{Diego} {$et~al$.}(2019){Diego}, {Hannuksela}, {Kelly}, {Pagano},
  {Broadhurst}, {Kim}, {Li}, \& {Smoot}}]{2019A&A...627A.130D}
{Diego}, J.~M., {Hannuksela}, O.~A., {Kelly}, P.~L., {$et~al$.} 2019, \aap,
  627, A130

\bibitem[{{Dimmelmeier} {$et~al$.}(2008){Dimmelmeier}, {Ott}, {Marek}, \&
  {Janka}}]{2008PhRvD..78f4056D}
{Dimmelmeier}, H., {Ott}, C.~D., {Marek}, A., \& {Janka}, H.~T. 2008, \prd, 78,
  064056

\bibitem[{{Edwards}(2021)}]{2021PhRvD.103b4025E}
{Edwards}, M.~C. 2021, \prd, 103, 024025

\bibitem[{Eker {$et~al$.}(2018)Eker, Bakış, Bilir, Soydugan, Steer, Soydugan,
  Bakış, Aliçavuş, Aslan, \& Alpsoy}]{Eker2018}
Eker, Z., Bakış, V., Bilir, S., {$et~al$.} 2018, Monthly Notices of the Royal
  Astronomical Society, 479, 5491–5511

\bibitem[{{Fryer} {$et~al$.}(2002){Fryer}, {Holz}, \&
  {Hughes}}]{2002ApJ...565..430F}
{Fryer}, C.~L., {Holz}, D.~E., \& {Hughes}, S.~A. 2002, \apj, 565, 430

\bibitem[{{Fryer} \& {New}(2011)}]{2011LRR....14....1F}
{Fryer}, C.~L., \& {New}, K. C.~B. 2011, Living Reviews in Relativity, 14, 1

\bibitem[{{Haris} {$et~al$.}(2018){Haris}, {Mehta}, {Kumar}, {Venumadhav}, \&
  {Ajith}}]{2018arXiv180707062H}
{Haris}, K., {Mehta}, A.~K., {Kumar}, S., {Venumadhav}, T., \& {Ajith}, P.
  2018, arXiv e-prints, arXiv:1807.07062

\bibitem[{{Kuroda} {$et~al$.}(2014){Kuroda}, {Takiwaki}, \&
  {Kotake}}]{2014PhRvD..89d4011K}
{Kuroda}, T., {Takiwaki}, T., \& {Kotake}, K. 2014, \prd, 89, 044011

\bibitem[{{Li} {$et~al$.}(2018){Li}, {Mao}, {Zhao}, \&
  {Lu}}]{2018MNRAS.476.2220L}
{Li}, S.-S., {Mao}, S., {Zhao}, Y., \& {Lu}, Y. 2018, \mnras, 476, 2220

\bibitem[{{Meena} \& {Bagla}(2020)}]{2020MNRAS.492.1127M}
{Meena}, A.~K., \& {Bagla}, J.~S. 2020, \mnras, 492, 1127

\bibitem[{{Mishra} {$et~al$.}(2021){Mishra}, {Meena}, {More}, {Bose}, \& {Singh
  Bagla}}]{2021arXiv210203946M}
{Mishra}, A., {Meena}, A.~K., {More}, A., {Bose}, S., \& {Singh Bagla}, J.
  2021, arXiv e-prints, arXiv:2102.03946

\bibitem[{{Moniez}(2010)}]{2010GReGr..42.2047M}
{Moniez}, M. 2010, General Relativity and Gravitation, 42, 2047

\bibitem[{{Morozova} {$et~al$.}(2018){Morozova}, {Radice}, {Burrows}, \&
  {Vartanyan}}]{2018ApJ...861...10M}
{Morozova}, V., {Radice}, D., {Burrows}, A., \& {Vartanyan}, D. 2018, \apj,
  861, 10

\bibitem[{{Mr{\'o}z} {$et~al$.}(2019){Mr{\'o}z}, {Udalski}, {Skowron},
  {Szyma{\'n}ski}, {Soszy{\'n}ski}, {Wyrzykowski}, {Pietrukowicz},
  {Koz{\l}owski}, {Poleski}, {Ulaczyk}, {Rybicki}, \&
  {Iwanek}}]{2019ApJS..244...29M}
{Mr{\'o}z}, P., {Udalski}, A., {Skowron}, J., {$et~al$.} 2019, \apjs, 244, 29

\bibitem[{{Mr{\'o}z} {$et~al$.}(2020){Mr{\'o}z}, {Udalski}, {Szyma{\'n}ski},
  {Soszy{\'n}ski}, {Pietrukowicz}, {Koz{\l}owski}, {Skowron}, {Poleski},
  {Ulaczyk}, {Gromadzki}, {Rybicki}, {Iwanek}, \&
  {Wrona}}]{2020ApJS..249...16M}
{Mr{\'o}z}, P., {Udalski}, A., {Szyma{\'n}ski}, M.~K., {$et~al$.} 2020, \apjs,
  249, 16

\bibitem[{{M{\"u}ller} {$et~al$.}(2013){M{\"u}ller}, {Janka}, \&
  {Marek}}]{2013ApJ...766...43M}
{M{\"u}ller}, B., {Janka}, H.-T., \& {Marek}, A. 2013, \apj, 766, 43

\bibitem[{{Nakamura} \& {Deguchi}(1999)}]{1999PThPS.133..137N}
{Nakamura}, T.~T., \& {Deguchi}, S. 1999, Progress of Theoretical Physics
  Supplement, 133, 137

\bibitem[{{Ohanian}(1974)}]{1974IJTP....9..425O}
{Ohanian}, H.~C. 1974, International Journal of Theoretical Physics, 9, 425

\bibitem[{{Ott}(2009)}]{2009CQGra..26f3001O}
{Ott}, C.~D. 2009, Classical and Quantum Gravity, 26, 063001

\bibitem[{{Ott} {$et~al$.}(2006){Ott}, {Burrows}, {Dessart}, \&
  {Livne}}]{2006PhRvL..96t1102O}
{Ott}, C.~D., {Burrows}, A., {Dessart}, L., \& {Livne}, E. 2006, \prl, 96,
  201102

\bibitem[{{Ott} {$et~al$.}(2013){Ott}, {Abdikamalov}, {M{\"o}sta}, {Haas},
  {Drasco}, {O'Connor}, {Reisswig}, {Meakin}, \&
  {Schnetter}}]{2013ApJ...768..115O}
{Ott}, C.~D., {Abdikamalov}, E., {M{\"o}sta}, P., {$et~al$.} 2013, \apj, 768,
  115

\bibitem[{{Peters}(1974)}]{1974PhRvD...9.2207P}
{Peters}, P.~C. 1974, \prd, 9, 2207

\bibitem[{{Piro} \& {Pfahl}(2007)}]{2007ApJ...658.1173P}
{Piro}, A.~L., \& {Pfahl}, E. 2007, \apj, 658, 1173

\bibitem[{{Powell} {$et~al$.}(2016){Powell}, {Gossan}, {Logue}, \&
  {Heng}}]{2016PhRvD..94l3012P}
{Powell}, J., {Gossan}, S.~E., {Logue}, J., \& {Heng}, I.~S. 2016, \prd, 94,
  123012

\bibitem[{{Powell} \& {M{\"u}ller}(2020)}]{2020MNRAS.494.4665P}
{Powell}, J., \& {M{\"u}ller}, B. 2020, \mnras, 494, 4665

\bibitem[{{Radice} {$et~al$.}(2019){Radice}, {Morozova}, {Burrows},
  {Vartanyan}, \& {Nagakura}}]{2019ApJ...876L...9R}
{Radice}, D., {Morozova}, V., {Burrows}, A., {Vartanyan}, D., \& {Nagakura}, H.
  2019, \apjl, 876, L9

\bibitem[{{Savitzky} \& {Golay}(1964)}]{1964AnaCh..36.1627S}
{Savitzky}, A., \& {Golay}, M.~J.~E. 1964, Analytical Chemistry, 36, 1627

\bibitem[{{Scheidegger} {$et~al$.}(2008){Scheidegger}, {Fischer}, {Whitehouse},
  \& {Liebend{\"o}rfer}}]{2008A&A...490..231S}
{Scheidegger}, S., {Fischer}, T., {Whitehouse}, S.~C., \& {Liebend{\"o}rfer},
  M. 2008, \aap, 490, 231

\bibitem[{{Scheidegger} {$et~al$.}(2010){Scheidegger}, {Whitehouse},
  {K{\"a}ppeli}, \& {Liebend{\"o}rfer}}]{2010CQGra..27k4101S}
{Scheidegger}, S., {Whitehouse}, S.~C., {K{\"a}ppeli}, R., \&
  {Liebend{\"o}rfer}, M. 2010, Classical and Quantum Gravity, 27, 114101

\bibitem[{{Schneider} {$et~al$.}(1992){Schneider}, {Ehlers}, \&
  {Falco}}]{1992grle.book.....S}
{Schneider}, P., {Ehlers}, J., \& {Falco}, E.~E. 1992, {Gravitational Lenses},
  doi:10.1007/978-3-662-03758-4

\bibitem[{{Somiya}(2012)}]{2012CQGra..29l4007S}
{Somiya}, K. 2012, Classical and Quantum Gravity, 29, 124007

\bibitem[{{Sotani} {$et~al$.}(2021){Sotani}, {Takiwaki}, \&
  {Togashi}}]{2021arXiv211003131S}
{Sotani}, H., {Takiwaki}, T., \& {Togashi}, H. 2021, arXiv e-prints,
  arXiv:2110.03131

\bibitem[{Suyama {$et~al$.}(2005)Suyama, Takahashi, \&
  Michikoshi}]{Suyama_2005}
Suyama, T., Takahashi, R., \& Michikoshi, S. 2005, Physical Review D, 72,
  doi:10.1103/physrevd.72.043001

\bibitem[{{Takahashi} \& {Nakamura}(2003)}]{2003ApJ...595.1039T}
{Takahashi}, R., \& {Nakamura}, T. 2003, \apj, 595, 1039

\bibitem[{{Unnikrishnan}(2013)}]{2013IJMPD..2241010U}
{Unnikrishnan}, C.~S. 2013, International Journal of Modern Physics D, 22,
  1341010

\bibitem[{{Vartanyan} \& {Burrows}(2020)}]{2020ApJ...901..108V}
{Vartanyan}, D., \& {Burrows}, A. 2020, \apj, 901, 108

\bibitem[{Vink(2020)}]{vink2020physics}
Vink, J. 2020, Physics and Evolution of Supernova Remnants (Springer)

\bibitem[{{Warren} {$et~al$.}(2020){Warren}, {Couch}, {O'Connor}, \&
  {Morozova}}]{2020ApJ...898..139W}
{Warren}, M.~L., {Couch}, S.~M., {O'Connor}, E.~P., \& {Morozova}, V. 2020,
  \apj, 898, 139

\bibitem[{{Wyrzykowski} {$et~al$.}(2011){Wyrzykowski}, {Skowron},
  {Koz{\l}owski}, {Udalski}, {Szyma{\'n}ski}, {Kubiak}, {Pietrzy{\'n}ski},
  {Soszy{\'n}ski}, {Szewczyk}, {Ulaczyk}, {Poleski}, \&
  {Tisserand}}]{2011MNRAS.416.2949W}
{Wyrzykowski}, L., {Skowron}, J., {Koz{\l}owski}, S., {$et~al$.} 2011, \mnras,
  416, 2949

\bibitem[{{Xu} {$et~al$.}(2021){Xu}, {Ezquiaga}, \&
  {Holz}}]{2021arXiv210514390X}
{Xu}, F., {Ezquiaga}, J.~M., \& {Holz}, D.~E. 2021, arXiv e-prints,
  arXiv:2105.14390

\bibitem[{{Yakunin} {$et~al$.}(2015){Yakunin}, {Mezzacappa}, {Marronetti},
  {Yoshida}, {Bruenn}, {Hix}, {Lentz}, {Bronson Messer}, {Harris}, {Endeve},
  {Blondin}, \& {Lingerfelt}}]{2015PhRvD..92h4040Y}
{Yakunin}, K.~N., {Mezzacappa}, A., {Marronetti}, P., {$et~al$.} 2015, \prd,
  92, 084040

\end{thebibliography}
\bibliographystyle{apj}

\appendix

\section{How Good is the Point-Mass Approximation?}
\label{sec:appendix_A}

A straightforward generalization to the point mass approximation is to model the microlens 
as a uniform-density sphere of radius $R_\star$. 
The corresponding three-dimensional gravitational potential is given as

\begin{figure*}[ht!]
	\centering
	\includegraphics[height=7.5cm, width=16.5cm]{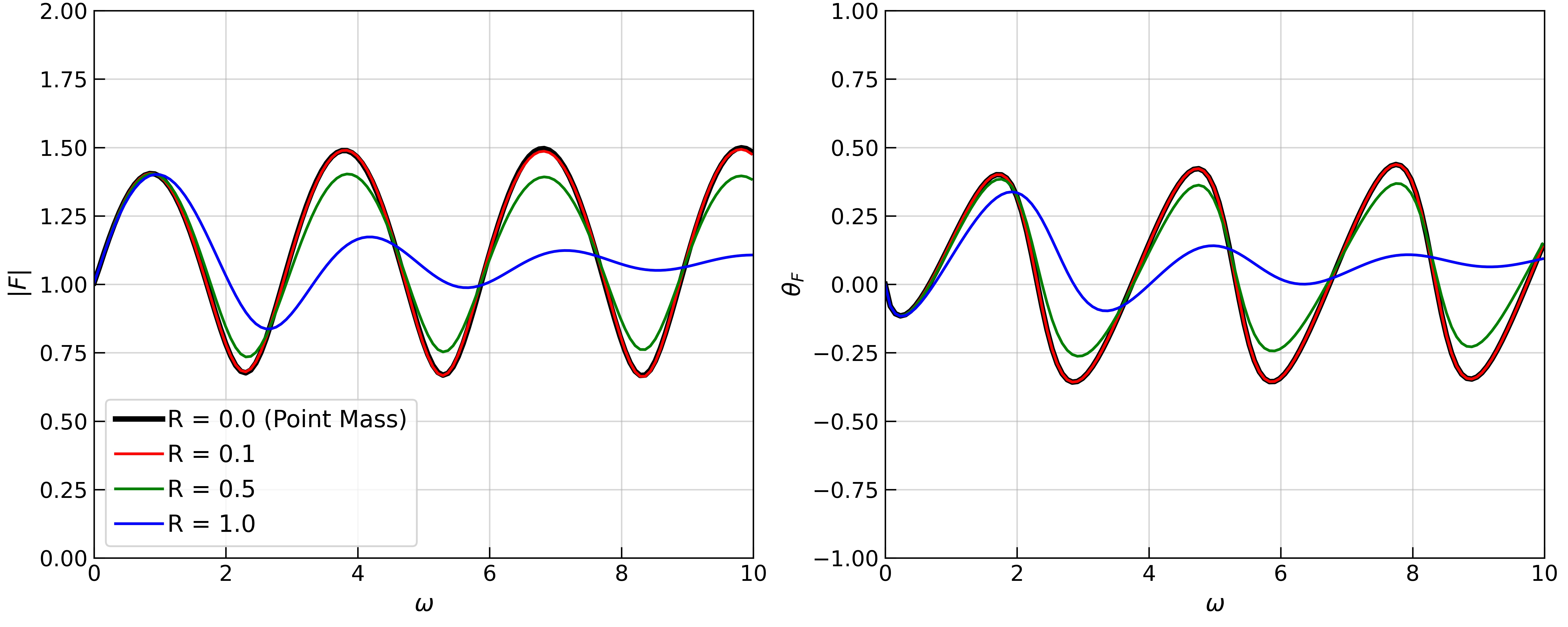}
	\caption{Comparison of uniform density sphere lenses to a
          point mass lens: absolute value and phase of the
          amplification factor as functions of dimensionless
          frequency, $\omega =   8 \pi G (1 + z_{\rm d}) {\rm M}_{\rm
            L} f / c^3$ are shown in left and right panel,
          respectively. The black line represents the point mass lens
          approximation. The colored lines represent the uniform
          density sphere approximation with different radii. When
          $R{\sim}0.1$ (or smaller), the two models produce almost
          identical results.  For $R{>}0.4-0.5$, the point mass
          approximation is no longer valid.} 
	\label{fig:pt_mass_apprx}
\end{figure*}

\begin{equation}
    \phi(r) = 
    \begin{cases}
        -\frac{GM}{2R_\star}(3 - \frac{r^2}{R_\star^2}) & ;r \leq R_\star,\\
        -\frac{GM}{R_\star} & ;r \geq R_\star,
    \end{cases}
\end{equation}
and the corresponding projected gravitational lensing potential is given as 
\citep{Suyama_2005}
\begin{equation}
    \psi(x) = 
    \begin{cases}
	    \ln[1 + (1 - (x/R)^2)^{0.5}]  \\
	    \quad - (1/3)[4 - (x/R)^2][1 - (x/R)]^{0.5} & ;x \leq R, \\
	    \ln[x/R] &      ;x \geq R,
    \end{cases}
\end{equation}
where $R {=} R_\star / \xi_0$ is the radius of the lens normalised by
the Einstein radius $\left({=}\sqrt{4 G M_{\rm L} D_{\rm d} D_{\rm ds}
  / c^2 D_{\rm s}}\right)$ of the Lens system.  
It turns out that this (dimensionless) parameter $R$ is what
determines the accuracy of the point mass approximation. 

In Figure~\ref{fig:pt_mass_apprx}, we shows the absolute and phase
values of the amplification factor for a point mass lens and uniform
density sphere lenses.  
When the radius of the lens is $R{\sim}0.1$ (or even smaller), the
point mass lens approximation leads to very accurate results.  
However, as the radius increases ($R\sim0.4-0.5$), we begin to notice
discrepancies between the two models at high frequency values. 
For even larger values of lens radius, the point mass lens
approximation no longer holds.    

To assess the physical situations during which the point mass
approximation holds, we next provide certain ballpark numbers for the
value of $R$:  

\begin{enumerate}
\item
  For a lens system in which all distances are of the order of one
  kpc, and for a solar mass lens, the Einstein radius is of the order
  of $10^{-9}$ rad. The angular size of the sun at this distance is
  ${\sim} 10^{-11}$ rad, and thus $R {\sim} 0.01$. Hence for typical
  galactic distances, a solar mass lens is well approximated by a
  point mass lens.  
\item
  For stars on the Main Sequence, $R_\star {\propto} M^{0.57}$
  \citep[e.g.][]{Eker2018}, while the Einstein radius ${\propto}
  M^{0.5}$. Thus, $R {\propto} M^{0.07}$. The value of $R$ remains
  fairly constant with increase in mass, and hence the point mass lens
  approximation holds true for even larger stars within typical
  galactic distances. 
\item
  When all relevant distances are of order $D$, the Einstein Radius
  ${\propto} \sqrt{D}$, while the angular size of an object ${\propto}
  D^{-1}$, and hence $R {\propto} \frac{1}{\sqrt{D}}$. At large values
  of D ($>$ 1kpc), the value of $R$ is bound to decrease. However, for
  values of $D < 1kpc$, the value of $R$ will begin to increase, and
  if $D$ happens to be a few tens of parsecs, $R$ will be of order
  unity. It is only in this unlikely scenario that the point mass
  approximation ceases to hold. 
\end{enumerate}

Given that the uniform sphere model is a highly simplified and
unrealistic model, one may  
try to use an alternate mass distribution profile. 
One possibility is that of a polytrope of index $n=3$, but as has been
shown in \citet{1974IJTP....9..425O}, the difference is again
noticeable only when $R$ is of order unity.

\end{document}